\documentclass[accepted]{uai2022} 

\usepackage[american]{babel}

\usepackage{natbib} 
\usepackage{mathtools} 
\usepackage{booktabs} 
\usepackage{tikz} 

\usepackage[utf8]{inputenc} 
\usepackage[T1]{fontenc}    
\usepackage{hyperref}       
\usepackage{url}            
\usepackage{booktabs}       
\usepackage{amsfonts}       
\usepackage{nicefrac}       
\usepackage{microtype}      
\usepackage{lipsum}		
\usepackage{graphicx}

\usepackage{setspace}
\usepackage{fullpage,graphicx,psfrag,amsmath,amsfonts,verbatim,tabularx,multirow,amssymb}
\usepackage{color,soul}
\usepackage{placeins}
\usepackage{tikz}
\usepackage{algorithm}
\usepackage[noend]{algpseudocode}
\usetikzlibrary{bayesnet}
\usetikzlibrary{arrows}
\usepackage{color}
\usepackage[justification=centering]{caption}
\usepackage{subcaption}
\newcommand{\multilinecomment}[1]{}
\usetikzlibrary{backgrounds}
\usepackage{amsthm}
\usepackage{hyperref}
\usepackage{natbib}      
\usepackage{booktabs}
\usepackage{mathtools}

\newcount\Comments  
\definecolor{darkgreen}{rgb}{0,0.5,0}
\newcommand{\kibitz}[2]{\ifnum\Comments=1\textcolor{#1}{#2}\fi}

\newcommand{\adigi}[1]{\kibitz{blue}{[AD: #1]}}

\newcommand{\slackmaximin}{\eta_{m}}
\newcommand{\slackexpl}{\eta_{e}}
\newcommand{\transienceN}{K}

\newtheorem{theorem}{Theorem}
\newtheorem{lemma}{Lemma}

\newtheorem{definition}{Definition}

\newtheorem{exmp}{Example}[section]
\DeclareMathOperator*{\argmax}{arg\,max}
\DeclareMathOperator*{\argmin}{arg\,min}

\makeatletter
\newtheorem*{rep@theorem}{\rep@title}
\newcommand{\newreptheorem}[2]{%
\newenvironment{rep#1}[1]{%
 \def\rep@title{#2 \ref{##1}}%
 \begin{rep@theorem}}%
 {\end{rep@theorem}}}
\makeatother
\newreptheorem{theorem}{Theorem}
\newreptheorem{lemma}{Lemma}
\newenvironment{sketch}{\paragraph{\normalfont \textit{Proof Sketch.}}}{\hfill$\square$}

\DeclarePairedDelimiter{\ceil}{\lceil}{\rceil}
\DeclarePairedDelimiter{\floor}{\lfloor}{\rfloor}

\usepackage{xr-hyper}
\makeatletter
\newcommand*{\addFileDependency}[1]{
  \typeout{(#1)}
  \@addtofilelist{#1}
  \IfFileExists{#1}{}{\typeout{No file #1.}}
}
\makeatother

\title{Balancing Adaptability and Non-exploitability in Repeated Games}

\author[1]{\href{mailto:<adigi@umich.edu>?Subject=Your UAI 2022 paper}{Anthony~DiGiovanni}{}}
\author[1]{Ambuj~Tewari}
\affil[1]{%
    Department of Statistics\\
    University of Michigan\\
    Ann Arbor, MI, USA
}
  
\begin{document}
\maketitle

\begin{abstract}
  We study the problem of adaptability in repeated games: simultaneously guaranteeing low regret for several classes of opponents.
We add the constraint that our algorithm is non-exploitable, in that the opponent lacks an incentive to use an algorithm against which we cannot achieve rewards exceeding some ``fair'' value.
Our solution is an expert algorithm (LAFF),
which searches within a set of sub-algorithms that are optimal for each opponent class,
and
punishes evidence of exploitation by switching to a
policy that enforces a fair solution.
With benchmarks that depend on the opponent class, we first show that LAFF has sublinear regret uniformly over 
these classes.
Second, we show that LAFF discourages exploitation, 
because exploitative opponents have linear regret.
To our knowledge, this work is the first to provide guarantees for both regret and non-exploitability in multi-agent learning.
\end{abstract}

\section{Introduction}\label{sec:intro}

General-sum repeated games
represent interactions between agents aiming to maximize their respective reward functions, with the possibility of compromise over conflicting goals. Despite their simplicity, achieving high rewards in such games is a challenging learning problem due to the complex space of 
possible opponents.
Both the behavior of a given opponent 
throughout
a game, and that opponent's choice of learning algorithm, may depend on one's own algorithm.
\citet{C20} 
argues,
based on empirical studies of repeated game tournaments, that a successful agent must achieve two goals. First, it must optimize its actions with respect to its beliefs about the opponent. Second, it should act such that
the opponent forms beliefs
motivating a response that is beneficial to the agent.

In particular, multi-agent reinforcement learning (MARL) features the following tradeoff: how to adapt to a variety of 
potential opponents,
while also actively shaping other agents' models of
oneself
such that they respond with cooperation, rather than exploitation.
If
an agent
commits to a
fixed policy
to ``lead'' the other player's best response \citep{LS01}, it may perform arbitrarily poorly against players that do not converge to such a response. This motivates the design of adaptive algorithms that try to lead,
but can
retreat
to a ``Follower'' (best response) approach if doing so gives greater rewards \citep{PS05, ICML10-chakraborty}.
An effective algorithm in this class is S++ \citep{C14}, which, 
due to its
Follower sub-algorithm, has the drawback that it is exploitable\textemdash that is, it rewards agents insisting on unfair bargains (``bully'' strategies) 
\citep{CO18, SLRC21}.

A simple motivating example of Follower exploitability is the game of Chicken (Figure \ref{fig:chicken}),
between players Row and Column. 
Suppose Column knows
Row
will take
the apparently optimal action 1 
if Column 
repeats action 2.
Column
will then want to use the Leader strategy of committing to action 2 to gain the highest reward. Row thus only gets reward 0.25, and if Column has truly committed, an attempt by Row to dissuade this strategy by taking action 2 would give both players reward 0.
A cooperative outcome, e.g., alternating between the off-diagonal cells, could be achieved if Row's learning algorithm were designed to \textit{publicly disincentivize} commitments
to the exploitative Leader strategy.

\begin{figure}[ht]
    \centering
    \begin{tabular}{|c|c|}
\hline
    0.5, 0.5 & 0.25, 1  \\
\hline
    1, 0.25 & 0, 0\\
\hline
\end{tabular}
    \caption{Reward bimatrix for Chicken.}
    \label{fig:chicken}
\end{figure}

MARL research has largely neglected the latter half of the adaptability vs. non-exploitability tradeoff.
Existing algorithms are either evaluated solely by 
their
rewards \textit{conditional} on given opponents \citep{PS05, C14}, or, when the evaluation criterion does account for the incentives of algorithm selection,
the pool of competitor algorithms typically excludes bully strategies \citep{CG10}.
Previous MARL algorithms addressing the adaptability half of the tradeoff lack finite-time guarantees on rewards.
We aim to provide a theoretically grounded algorithm for repeated games that is both adaptable, by using Leader and Follower sub-algorithms, and non-exploitable.
More broadly,
this paper addresses a challenge of interest in several
areas of machine learning:
designing algorithms that account for how the distribution of data the algorithms are applied to may change based on the choice of the algorithms themselves.

\paragraph{\textbf{Related work}} 
Previous algorithms for repeated games have
combined Leader and Follower modules,
aiming for
the following guarantees: worst-case safety, best response to players with bounded memory, and convergence in self-play to Pareto efficiency, i.e., an outcome in which no player can do better without the other doing worse \citep{PS04}.
Like ours,
these algorithms aim for adaptability,
but they do not have regret guarantees --- the desired
properties are only
shown to hold asymptotically.
Manipulator \citep{PS05} achieves these properties by starting with a fixed strategy
that maximizes the user's rewards conditional on the opponent using a best response, and switching to
reinforcement learning (RL) with a safety override if
that
strategy does not yield its target rewards.
Related to the self-play guarantee,
we prove a more general property of Pareto efficiency against effective RL algorithms (see Section \ref{sec:sub:rgclass}).
Like Manipulator, our approach tests
sub-algorithms sequentially.
S++ \citep{C14} 
has empirically strong performance
on
the guarantees above.
However,
neither of these algorithms guarantee non-exploitability.

Although to our knowledge
no previous works have proven non-exploitability in our sense,
several algorithms are designed to
achieve ``fair'' Pareto efficiency
in self-play without using
Follower approaches that would be exploitable.
\citet{LS05}'s algorithm for 
computation of
Nash equilibria, like our Leader sub-algorithms, enforces a Pareto efficient outcome 
by punishing deviations.
If an agent played this equilibrium, which satisfies properties of symmetry similar to
the outcome our Egalitarian Leader sub-algorithm aims for, it would be non-exploitable.
However, committing to this equilibrium
precludes
learning a best response to fixed strategies that offer higher rewards than the cooperative solution, or exploiting adaptive players, which our Conditional Follower and Bully Leader sub-algorithms achieve, respectively.
In two-player bandit problems where the reward bimatrix must be learned, UCRG \citep{TD20} has
near-optimal
regret in self-play with respect to the egalitarian bargaining solution
(Section \ref{bargtheory}).
However, it cannot provably cooperate with 
agents other than
itself, learn best responses, or exploit adaptive players.

Our objectives of adaptability and non-exploitability are inspired by work on learning equilibrium \citep{BT04, fcl, CR21}, a solution concept in which players' \textit{learning algorithms} are in a Nash equilibrium, beyond merely the equilibrium of an individual game itself.
This objective accounts for the dependence of the problems faced by multi-agent learning algorithms on the design of such algorithms. 

\paragraph{\textbf{Contributions}} We propose an algorithm (LAFF) that, to our knowledge, is the first proven to have both strong performance against different classes of players in repeated games and a guarantee of non-exploitability, formalized in Section \ref{sec:sub:regretdef}. Specifically, these classes consist of stationary
algorithms (``Bounded Memory''), unpredictable adversaries (``Adversarial''), and adaptive RL agents (``Follower''). 
LAFF's modular design
allows for extensions to a broader variety of opponent classes in future work. We propose regret metrics appropriate for games against Followers, based on the goal of Pareto efficiency. Our method of proof of adaptability and non-exploitability is novel, applying ``optimistic'' principles at two levels. First, LAFF starts with the sub-algorithm (or \textit{expert}) that would give the highest expected rewards 
if the opponent were
in that expert's target class (``potential''), then proceeds through experts in descending order of 
potential.
Second, LAFF chooses whether to switch experts by comparing the potential 
of the active expert with its empirical average reward plus a slack term, which decreases with the time for which the expert is used.
For non-exploitability and regret against Followers, we use the properties of an enforceable bargaining solution (see Section \ref{bargtheory}) to upper-bound the other player's rewards.

\section{Preliminaries}\label{sec:prelim}

We study a special class of Markov games: repeated games with a bounded memory state representation \citep{PS05} and public randomization.

\subsection{Setup and Opponent Classification}\label{sec:sub:rgclass}

\noindent Consider a repeated game over $T$ time steps, defined for players $i=1,2$ by action spaces $\mathcal{A}^{(i)}$,
reward matrices $\mathbf{R}^{(i)}$,
and a fixed player memory length $K \in \mathbb{N}$. Here, all $\mathbf{R}^{(i)}(a^{(1)}, a^{(2)}) \in [0,1]$ are known by both players.
At time~$t$ the following random variables are drawn: $S_t$ for state, $A_t^{(i)}$ for actions, and $R_t^{(i)} = \mathbf{R}^{(i)}(A_t^{(1)},A_t^{(2)})$ for rewards.
A state space $\mathcal{S} := (\mathcal{A}^{(1)})^K \times (\mathcal{A}^{(2)})^K \times \{0, 1\}^{2K+2}$, and transition probabilities $\mathcal{P}(s'|s,a^{(1)},a^{(2)})$ between states, are induced by two features:
(1) the tuple of both players' last $K$ actions, and (2) the tuple of the last $K$ and current outcome of a randomization signal, for each player. (See Section 2.1.2 of \citet{MS06}.)
Thus, players condition their 
actions
on their memory of the last $K$ time steps,
and a signal that permits
correlated action choices.

Formally, let $(w^{(1)}_t, w^{(2)}_t) \in [0, 1]^2$ be weights chosen by the respective players at time $t$,\footnote{We restrict to cases where players commit to a fixed weight, so the effective action space is finite. See the Appendix for details.} and draw $X_t \sim \text{Unif}[0,1]$ independent of all other random variables in the game. Then, letting $y_t^{(i)}$ be the realized value of $Y_t^{(i)} := \mathbb{I}[X_t < w^{(i)}_t]$, the second feature at time $t$ is
$(y_{t-K}^{(1)},...,y_t^{(1)},y_{t-K}^{(2)},...,y_t^{(2)})$.
This allows the players to correlate actions through the public signal $X_t$, even if one player unilaterally generates the signal.
For instance, in Chicken (Figure \ref{fig:chicken}),
players could flip a fair coin ($w^{(1)}_t = w^{(2)}_t = 0.5$)
at
each time step
and play the pair of actions
leading to the top-right cell
when it comes up heads, otherwise
play the bottom-left cell.
In this framework, at each time step each player has a choice of both a weight $w_t^{(i)}$ and policy $\pi^{(i)}_t: \mathcal{S} \to \Delta^{|\mathcal{A}^{(i)}|}$, a mapping from states to distributions over actions.

Given a fixed policy of player 2, a repeated game is a
Markov decision process (MDP) given by
$(\mathcal{S}, \mathcal{A}^{(1)}, r, p)$
as follows.
Let $a^{(i)}(s)$ be the last action of player $i$
that defines state $s$.
Here, $r: \mathcal{S} \times \mathcal{A}^{(1)} \to [0,1]$ is
$r(s, a) = \mathbf{R}^{(1)}(a^{(1)}(s), a^{(2)}(s))$,
and $p:\mathcal{S} \times \mathcal{A}^{(1)} \times \mathcal{S} \to [0,1]$ is
$p(s'|s,a) = \sum_{a^{(2)}} \mathcal{P}(s'|s,a,a^{(2)}) \pi^{(2)}(a^{(2)}|s)$.
A policy is called Markov if it is conditioned only on the current state.

The problem faced by our learner, player 1, depends on which of the following classes player 2's algorithm is in:
\begin{enumerate}
    \item \textit{Bounded Memory}: (i) Player 2 uses a constant $w^{(2)}$, reported at the start of the game; (ii) $\pi^{(2)}$ is Markov
    and does not depend on time or player 1's signals $w^{(1)}_t$ or $y_t^{(1)}$; and (iii) for all $s, a^{(2)}$ we have $\pi^{(2)}(a^{(2)}|s) > 0$.\footnote{This relatively strong condition is needed for a concentration result in our analysis, ruling out cases where players remain in a transient state for an unknown time. We need to know the exit time from the transient states to compute the quantity $\overline{r}_{i,
    \tau}^{(2)}$ used by one of our experts. Section \ref{sec:experiments} shows strong results against a Bounded Memory player (FTFT) for which this condition does not hold.} 
    \item \textit{Adversarial}: Player 2 selects actions according to any arbitrary distribution, which may depend on the history of play and on player 1's policy at each time step.
    \item \textit{Follower}: A Follower learns a best response when player 1 is ``eventually stationary'' (formalizing the follower concept in \citet{LS01}), and when the value of that best response meets player 2's standard of fairness. For some fairness threshold $V^{(2)} \geq 0$ (depending on the game), player 2's algorithm has the following properties.
    Suppose that after time $T_0$, player 1 always plays a Bounded Memory algorithm (without condition 3), which induces an MDP of finite diameter $D$ where player 2's optimal average reward is at least $V^{(2)}$.
    Then with probability at least $1-\delta$, player 2's regret up to time $T$ (see Section \ref{sec:sub:regretdef}) is bounded by $C_1T_0 + C_2D(SAT\log(T/\delta))^{1/2}$ for constants $C_1, C_2$. 
\end{enumerate}

A repeated game against a Bounded Memory player is equivalent to a communicating MDP \citep{puterman}.
A Follower formalizes an agent that models \textit{our} agent as an MDP (Leader), and the regret bound in our definition is of a standard form for RL algorithms \citep{optQ}. 
Many MARL algorithms take this approach at least partly \citep{PS05, ICML10-chakraborty, CG10}, hence this is a reasonable class to consider.
For example,
\citet{LS05}'s algorithm,
which plays a 
certain
sequence of actions
and punishes deviations from that sequence,
is Bounded Memory ---
this algorithm does not change its policy
in response to the other player,
but its policy conditions on past actions.
A standard RL algorithm,
which would learn the sequence played by \citet{LS05}'s algorithm
and converge to
an optimal policy against it,
and which is a component of more complex repeated games
algorithms like Manipulator and S++,
is a case of a Follower.

As discussed in \citet{C20},
a large proportion of top-performing algorithms are Bounded Memory (Leaders) or Followers, or switch between the two.
These classes
illustrate fundamental
approaches to multi-agent learning
(thus, likely opponents
that our algorithm would face):
Either an agent behaves consistently, trying to shape the learning opponent’s behavior (Bounded Memory), or 
the agent changes policies in a process of learning how the opponent behaves and computing an optimal response to that opponent, possibly subject to fairness standards as they try to avoid exploitation (Follower).
The Adversarial class accounts for opponent behavior between these two extremes, which is difficult to learn in generality, but a
worst-case guarantee
can still be achieved.
We thus restrict to guarantees against formalizations of these classes.
Bounds against a wider variety of opponents would be less theoretically tractable, as far as finding the optimal strategy against one class interferes with performance against another.
(For example, \citet{PS05} note that in the repeated Prisoner's Dilemma,
it is
impossible for an algorithm to guarantee the best
response to an opponent
that may play either grim trigger
---
``defect if and only if either
player defected last round''
---
or ``always cooperate.'')
Extending to other opponent classes is an important direction for future work.

\subsection{Background on Bargaining Theory}\label{bargtheory}

\noindent To define appropriate 
optimality criteria
for these opponent classes and construct corresponding experts, we use several concepts from bargaining theory.
We also illustrate these
concepts in the game of Chicken
from the introduction
(Example \ref{example:barg_concepts}).
Define the \textit{security values} 
$\mu_{\textsc{S}}^{(i)} := \max_{\mathbf{v}_i} \min_{\mathbf{v}_{-i}} \mathbf{v}_1^\intercal \mathbf{R}^{(i)} \mathbf{v}_2$,
i.e., the rewards that each player can guarantee
regardless of their opponent's actions,
with player 1's maximin strategy as $\mathbf{v}^{(1)}_{\textsc{M}} = \argmax_{\mathbf{v}_1} \min_{\mathbf{v}_2} \mathbf{v}_1^\intercal \mathbf{R}^{(1)} \mathbf{v}_2$.
Let $\mathcal{G} := \{(\mathbf{R}^{(1)}(i,j),$ $\mathbf{R}^{(2)}(i,j)) \ | \ i \in \mathcal{A}^{(1)}, j \in \mathcal{A}^{(2)}\}$,
the set of reward pairs achievable
by pure actions in the game.
An important set of rewards in the computation of enforceable bargaining solutions is the convex polytope $\mathcal{U} := \text{Conv}(\mathcal{G}) \cap \{(u_1, u_2) \ | \ u_1 \geq \mu_{\textsc{S}}^{(1)}, u_2 \geq \mu_{\textsc{S}}^{(2)}\}$,
reward pairs that are achievable by randomizing over joint actions and give each player at least their security value.
One reward pair satisfying several desirable properties is the egalitarian bargaining solution (EBS) \citep{TD20}, given by $(\mu_{\textsc{E}}^{(1)},  \mu_{\textsc{E}}^{(2)}) := \argmax_{(u_1, u_2) \in \mathcal{U}} \min_{i=1,2}\{u_i - \mu_{\textsc{S}}^{(i)}\}$.

The reward pairs over which we search for optimal benchmark values,
described in Section \ref{sec:sub:regretdef},
are subject to the following constraint of enforceability. To our knowledge, this definition, including the formalization of enforceability for finite punishment lengths, has not been provided in previous work on non-discounted games. However, see Definition 2.5.1 in \citet{MS06} for the discounted case.

\begin{definition}
\label{def:enf}
Let $(u_1, u_2) \in \mathcal{U}$ be a convex combination
of points in some set of joint actions $\mathcal{X}$.
Let $r(\mathcal{X}) := \max_{(x_1,x_2) \in \mathcal{X}} \{\max_{j \neq x_2} \mathbf{R}^{(2)}(x_1,j) - \mathbf{R}^{(2)}(x_1,x_2)\}$
be player 2's deviation profit.
Then $(u_1, u_2)$ is \textbf{$\epsilon$-enforceable}, relative to a memory length $K$ and $\epsilon > 0$, if:
\begin{align*}
    Ku_2 &\geq K\mu_{\textsc{S}}^{(2)} + r(\mathcal{X}) + \epsilon.
\end{align*}
\end{definition}

Intuitively, if player 2 does not deviate from player 1's desired action sequence, player 2 receives
$u_2$
on average
for each of $K$ steps. If player 2 deviates, gaining at most $r(\mathcal{X})$ profit, player 1 may punish with player 2's security value for $K$ steps. We call the total sequence reward ``enforceable'' if 
it exceeds the total deviation reward
by at least $\epsilon$.
Let $\mathcal{U}(\epsilon)$ be the set of $\epsilon$-enforceable rewards in $\mathcal{U}$. Then, the feasible region $\mathcal{U}(\epsilon)$, 
used to compute an enforceable version of the EBS,
shrinks with increasing~$\epsilon$ and decreasing~$K$.

The $\epsilon$-enforceable EBS, which we will use to design one of the Leader experts, is found by solving the optimization problem from Section 3.2.4 of \citet{TD20} under the constraint in Definition \ref{def:enf}.
A similar procedure, applied to the objective of maximizing only player 1's reward, gives the Bully solution for the second
Leader expert.
We provide details on these solutions in the Appendix.

\begin{exmp}
\label{example:barg_concepts}
In Chicken (Figure \ref{fig:chicken}),
both players' security value is 0.25, guaranteed by playing action 1.
The EBS is given by 50\% weight on the top-right action pair, and 50\% on the bottom-left, giving both players $0.625$.
If player~1 plays its half of either action pair in the EBS, player 2 does worse by deviating
(by a margin of at least 0.25), so no punishment
is necessary to enforce
the EBS.
Thus the EBS is enforceable for any $K$ and $\epsilon < 0.375K + 0.25$.
\end{exmp}

\subsection{Objectives}\label{sec:sub:regretdef}

\noindent The metric of regret, which we aim to minimize, varies based on the class of player 2 our algorithm faces. For a player 2 algorithm $\mathfrak{B}$, regret with respect to a benchmark $\mu(\mathfrak{B})$ is $\mathcal{R}(T) := T\mu(\mathfrak{B}) - \sum_{t=1}^T R_t^{(1)}$.

\paragraph{\textbf{Bounded Memory}} By condition 3 for Bounded Memory, player 2 induces a communicating MDP.
Let $\Pi$ be the set of time-independent deterministic Markov policies. Then the state-independent optimal average reward is $\mu_*^{(1)} := \max_{\pi^{(1)} \in \Pi} \lim_{t \to \infty} \frac{1}{t} \mathbb{E}_{ \pi^{(1)}}(\sum_{i=0}^t R_i^{(1)}|S_0)$. Here, $\mu(\mathfrak{B}) = \mu_*^{(1)}$.

\paragraph{\textbf{Adversarial}} Against an Adversarial player, an appropriate benchmark is the greatest expected value that player 1 can guarantee, no matter player 2's actions. This is player~1's security value: $\mu(\mathfrak{B}) = \mu_{\textsc{S}}^{(1)}$. Note the distinction from \textit{external regret} used in adversarial bandits and MDPs.
While the problem is trivial if player 2 is known to be Adversarial, since one can always play the maximin strategy, our challenge is to maintain low Adversarial regret without losing guarantees on other regret measures. This corresponds to \textit{safety} in multi-agent learning \citep{PS04}.

\paragraph{\textbf{Follower}} The concept of regret against a Follower is more complex.
Player 2's sequence of policies can vary significantly based on
player 1's actions. 
Evaluating our algorithm 
by
the maximum average reward in hindsight would have to account for this counterfactual dependence \citep{C14}.
However, by considering enforceability, we can define benchmarks by lower bounds on this maximum,
constrained
by the Follower's fairness value $V^{(2)}$.
We consider two cases depending on $V^{(2)}$, focusing for simplicity on the extremes where the Follower either accepts nothing less than the EBS or accepts any enforceable bargain. In principle, our framework could be extended for other $V^{(2)}$ values.

First, the EBS
is Pareto efficient, meaning 
we cannot achieve greater than $\mu_{\textsc{E}}^{(1)}$ without player 2 receiving less than $\mu_{\textsc{E}}^{(2)}$.
When
the EBS can be enforced
with a fixed policy, $\mu_{\textsc{E}}^{(1)}$ is thus an appropriate 
benchmark if the fairness threshold $V^{(2)}$ is player 2's part of the EBS pair.
The EBS is not always enforceable for finite $K$, however. 
In this case, 
the enforceable version of the
EBS is 
the maximizer
$(\mu_{\textsc{E},\epsilon}^{(1)}, \mu_{\textsc{E},\epsilon}^{(2)})$ of the objective $f(u_1, u_2) = \min_{i=1,2}\{u_i - \mu_{\textsc{S}}^{(i)}\}$ in $\mathcal{U}(\epsilon)$ for some $\epsilon > 0$. 
For this first case,
we therefore consider $V^{(2)} = \mu_{\textsc{E},\epsilon}^{(2)}$, where player 2 follows conditionally. If $\mathcal{U}(\epsilon)$ is empty, $(\mu_{\textsc{E},\epsilon}^{(1)}, \mu_{\textsc{E},\epsilon}^{(2)}) := (\mu_{\textsc{S}}^{(1)}, \mu_{\textsc{S}}^{(2)})$. We set $\mu(\mathfrak{B}) = \mu_{\textsc{E},\epsilon}^{(1)}$.

The second case is $V^{(2)} = 0$, i.e., player 2 follows unconditionally. Here, we compute the maximizer over $\mathcal{U}(\epsilon)$ of $f(u_1, u_2) = u_1$.
Let $(\mu_{\textsc{B},\epsilon}^{(1)}, \mu_{\textsc{B},\epsilon}^{(2)})$ be the solution to this optimization problem (the \textit{Bully values}), or $(\mu_{\textsc{B},\epsilon}^{(1)}, \mu_{\textsc{B},\epsilon}^{(2)}) := (\mu_{\textsc{S}}^{(1)}, \mu_{\textsc{S}}^{(2)})$ if no solution exists. We define $\mu(\mathfrak{B}) = \mu_{\textsc{B},\epsilon}^{(1)}$.

While these regret metrics 
provide standards for
adaptability,
we must also formalize non-exploitability. 
We seek a guarantee on an algorithm's performance against its best response.
It is unclear how to characterize the best response to an algorithm capable of adapting to several opponent classes. Given this, we focus on a tractable and practically relevant subproblem: guaranteeing that the best response to our algorithm is not a ``bully'' in the sense discussed in the introduction, which is the most common exploitative strategy in MARL literature \citep{PS05, LS01,Press10409, LS05}.
Even this weaker guarantee is absent from previous work, and we show numerically in Section \ref{sec:experiments} that this suffices for our algorithm to be in learning equilibrium with itself
(see Section \ref{sec:intro}) in a pool of top-performing algorithms.

\begin{definition}
Let player 2 be Bounded Memory, 
and $\mu_{\textsc{M}}^{(1)}$ and $\mu_{\textsc{M}}^{(2)}$ be the expected rewards for players 1 and 2 
when player 1 uses $\mathbf{v}^{(1)}_{\textsc{M}}$ and
player 2 uses $\pi^{(2)}$.
An algorithm $\mathfrak{A}$ is
\textbf{$(V^{(1)},\slackexpl)$-non-exploitable} 
if, whenever
$\mu_*^{(1)} < V^{(1)} - \slackexpl$ and $\mu_{\textsc{M}}^{(2)} > \mu_{\textsc{E},\epsilon}^{(2)}$, for all $c > 0$ player 2's regret with respect to $\mu_{\textsc{E},\epsilon}^{(2)} + c$ against $\mathfrak{A}$ is $\Omega(T)$.
\end{definition}

Our algorithm is exploitable if player 2 can profit
(do better than $\mu_{\textsc{E},\epsilon}^{(2)}$)
from
a policy against which we cannot achieve close to
some value corresponding to a standard of fairness.
The hyperparameter $V^{(1)}$ tunes the tradeoff
between exploitability and
flexibility to various opponents.
Player 2 does \textit{not} profit from
exploitation if they incur linear regret.

\begin{exmp}
In Chicken (Figure \ref{fig:chicken}), let $V^{(1)} = 0.625$ (i.e., the EBS), and consider the following strategies: a) always play action 2, b) always play the opponent's last action,
and c) play the best response to the empirical distribution of the opponent's past actions. Strategy (a) is exploitative Bounded Memory. Thus, we argue that an effective algorithm should avoid playing the ``best response'' of action 1, instead discouraging the use of this strategy by, e.g., consistently playing the EBS (see Egalitarian Leader in the next section). Strategy (b) is also Bounded Memory, but not exploitative since one can achieve at least $V^{(1)}$ against this player on average. Our algorithm should therefore learn the best response to (b). Strategy (c) is a Follower with $V^{(2)} = 0$, thus our algorithm should converge to consistently playing action 2 against (c), achieving the Bully value.
\end{exmp}

\section{Lead and Follow Fairly (LAFF)}\label{sec:ergalgo}

We apply an expert algorithm to a set of experts designed for our target classes. Expert algorithms 
use an active expert to choose an action at a given time,
and switch active experts based on their relative performance \citep{C14}.
LAFF switches experts sequentially, going to the next expert in a predefined sequence only
if the rewards obtained by its active expert fall short of the current target value.
Some of the experts are also designed to guarantee non-exploitability.

\subsection{Description of Experts}

\begin{figure}
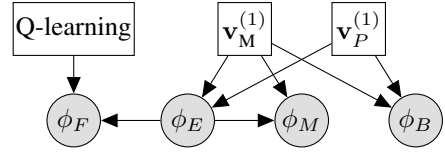

\centering
  \tikz{
 \node[obs, xshift=-3.5cm] (f) {$\phi_F$}; %
 \node[obs, xshift=-2cm] (e) {$\phi_E$}; %
 \node[obs, xshift=-0.5cm] (m) {$\phi_M$}; %
 \node[obs, xshift=1cm] (b) {$\phi_B$}; %
 \node[latent, rectangle, above=of f, yshift=-0.5cm] (q) {Q-learning};
 \node[latent, rectangle, above=of e, yshift=-0.5cm, xshift=0.75cm] (v) {$\mathbf{v}^{(1)}_{\textsc{M}}$};
 \node[latent, rectangle, above=of e, yshift=-0.5cm, xshift=2.25cm] (p) {$\mathbf{v}^{(1)}_P$};
 \edge {q} {f}
 \edge {v} {e,m,b}
 \edge {p} {e,b}
 \edge {e} {f,m}
 }
\caption{Algorithmic components (white) of LAFF's experts (gray). An arrow from one node to another means the former is used in computation of the output by the latter.}
\label{flowchart}
\end{figure}

\noindent 
LAFF uses an active expert for an epoch of length $H$ before checking whether to switch. Let $\tau$ be the time elapsed since LAFF started using the current instance of the active expert (at time $t_i + 1$), and define $\overline{r}^{(1)}_{i,\tau} := \frac{1}{\tau} \sum_{t=t_i + 1}^{t_i + \tau} R^{(1)}_t$ and $\overline{r}^{(2)}_{i,\tau} := \frac{1}{\tau-K} \sum_{t=t_i + K + 1}^{t_i + \tau} R^{(2)}_t$. See Figure \ref{flowchart} for a summary of algorithmic elements that these experts depend on.

\paragraph{\textbf{Conditional Follower $(\phi_F)$}}
Recall the
benchmarks $\mu_{\textsc{B},\epsilon}^{(1)}$,
$\mu_{\textsc{E},\epsilon}^{(1)}$, and
$\mu_{\textsc{S}}^{(1)}$from
Section \ref{sec:sub:regretdef}.
To handle cases where
$\mu_*^{(1)}$
against a Bounded Memory player 2
lies
between these values,
LAFF uses $\phi_F$ multiple times in the sequence (called ``instances''). This expert starts off equivalent to Optimistic Q-learning \citep{optQ}, whose regret bound
(in an MDP with $S$ states and $A$ actions)
with probability at least $1-\delta$ is $\mathcal{R}_{Q}(\tau, \delta) = \mathcal{O}((SA\log(\frac{\tau}{\delta}))^{1/3}\tau^{2/3})$. After each \textit{subepoch} of length $H^{1/2}$, if $\overline{r}^{(1)}_{i,\tau} < V^{(1)} - \frac{\mathcal{R}_{Q}(\tau, \delta/T)}{\tau}$, this expert switches to the Egalitarian Leader $\phi_E$ (below) for as long as \textit{any} instance of $\phi_F$ is used. Otherwise, it uses Optimistic Q-learning for the next subepoch.

\paragraph{\textbf{Conditional Maximin ($\phi_M$)}}
Initially, $\phi_M$ uses the policy $\pi^{(1)}(\cdot|s) = \mathbf{v}^{(1)}_{\textsc{M}}$ for all $s$. Let $\slackmaximin > 0$ be a slack variable, chosen based on the class of Adversarial players considered in Theorem \ref{hedge}. After each subepoch, if $\overline{r}^{(2)}_{i,\tau} > \mu_{\textsc{E},\epsilon}^{(2)} - \slackmaximin + \sqrt{\frac{\log(T/\delta)}{2(\tau-K)}}$, this expert switches to $\phi_E$ for the rest of the game. Otherwise, it uses $\mathbf{v}^{(1)}_{\textsc{M}}$ for the next subepoch.

\paragraph{\textbf{Egalitarian Leader ($\phi_E$)}} If there is no enforceable EBS, let $\phi_E \equiv \mathbf{v}^{(1)}_{\textsc{M}}$.
Otherwise, let the EBS action pairs be denoted $(a_{\textsc{E}}^{(1)}(y), a_{\textsc{E}}^{(2)}(y))$ for $y=0,1$,
and the weight on the first action pair
be $\alpha_{\textsc{E}}$.
While $\epsilon$-enforceability requires that a punishment of length $K$ is sufficient to make a reward pair player 2's best response, this length may not be \textit{necessary}.
We therefore consider the least harsh punishment (if any) needed to enforce the EBS, that is, the value $K' \leq K$ satisfying $K' = \max\Big\{0, \Big \lceil \frac{r(\{(a_{\textsc{E}}^{(1)}(0), a_{\textsc{E}}^{(2)}(0)), (a_{\textsc{E}}^{(1)}(1), a_{\textsc{E}}^{(2)}(1))\}) + \epsilon}{\mu_{\textsc{E},\epsilon}^{(2)} - \mu_{\textsc{S}}^{(2)}} \Big \rceil \Big\}$.

Let $\mathbf{v}^{(1)}_P := \argmin_{\mathbf{v}_1} \max_{\mathbf{v}_2} \mathbf{v}_1^\intercal\mathbf{R}^{(2)}\mathbf{v}_2$, player 1's punishment strategy.
Recall that policies in our framework are conditioned on binary signals $Y_t^{(i)}$,
whose distributions are determined
by players' reported weights $w_t^{(i)}$.
Then, for the first ${K'}$ time steps, with the realized value $y_{t}^{(1)}$ of the signal given by $w_t^{(1)} = \alpha_{\textsc{E}}$ for all $t$, $\phi_E$ plays $a_{\textsc{E}}^{(1)}(y_{t}^{(1)})$. (This
ensures that, if LAFF switches to $\phi_E$ mid-game, player 2 is not punished for
having played actions other than the EBS
before LAFF started signaling enforcement of the EBS.) Afterwards, $\phi_E$ uses the following stationary policy. If, for any of the past $K'$ timesteps, player 2 has played $A^{(2)}_t \neq a_{\textsc{E}}^{(2)}(y_t^{(2)})$
--- i.e., deviated from the EBS ---
the distribution over actions for that state is $\mathbf{v}^{(1)}_P$. Otherwise, $a_{\textsc{E}}^{(1)}(y_t^{(1)})$ is played.

\paragraph{\textbf{Bully Leader ($\phi_B$)}} This expert is defined like $\phi_E$, but using the Bully solution from Section \ref{bargtheory}
(maximizing the selfish objective).
If there is no enforceable solution, given by $(a_{\textsc{B}}^{(1)}(y), a_{\textsc{B}}^{(2)}(y))$ for $y=0,1$ and $\alpha_{\textsc{B}}$, let $\phi_B \equiv \mathbf{v}^{(1)}_{\textsc{M}}$. Otherwise, define $\phi_B$ just as $\phi_E$ for this solution.

\subsection{Algorithm}

We design the selection of experts by LAFF (Algorithm \ref{followfirst}) such that, for any of our target classes, LAFF eventually
commits to the optimal expert against player 2 in a sequence $\{\phi_j\}_j$.
Over an epoch, the active expert is executed,
and we update this expert's average rewards
since it was made active (line \ref{record}). Afterwards, LAFF switches to the next expert in the schedule if and only if it rejects the hypothesis that the current expert's expected value exceeds its corresponding target $\mu_j$ (line \ref{baselinecheck}).
The false positive rate of this hypothesis test is controlled by a function $\mathcal{B}$, which
decreases with 
$\sqrt{\tau}$.
We define $\mathcal{B}$ in the proof of Lemma \ref{followregret} (see Appendix).
\multilinecomment{
The false positive rate of this hypothesis test is controlled by a function $\mathcal{B}$ of the time elapsed since the last switch (line \ref{tauup}), defined:
\begin{align*}
    \xi(\epsilon, r) &:= \begin{cases}
    \frac{\epsilon}{2K'},& \text{if } r \geq 0\\
    \frac{\epsilon + r}{2K'},& \text{if } -\epsilon < r < 0\\
    -r,              & \text{otherwise},
\end{cases} \\
    \mathcal{B}(\tau) &:= \frac{1}{\tau} \cdot \frac{K'\xi(\epsilon, r(\mathcal{X})) + C_1T_0 + K'+1}{\xi(\epsilon, r(\mathcal{X}))} \\
    &+ \frac{1}{\tau} \cdot \frac{C_2\mathcal{R}_{Q}(\tau, \frac{\delta}{T}) + (3 + \xi(\epsilon, r(\mathcal{X})))\sqrt{\frac{\tau \log(\frac{T}{\delta})}{2}}}{\xi(\epsilon, r(\mathcal{X}))}.
\end{align*}
Where $\mathcal{X} = \mathcal{X}_{\textsc{B}} := \{(a_{\textsc{B}}^{(1)}(y), a_{\textsc{B}}^{(2)}(y))\}_{y=0,1}$ for expert index $j \leq 2$, $\mathcal{X} = \mathcal{X}_{\textsc{E}} := \{(a_{\textsc{E}}^{(1)}(y), a_{\textsc{E}}^{(2)}(y))\}_{y=0,1}$ for $j > 2$, and $\delta > 0$ is some confidence level.
}
Because $\mu_{\textsc{B},\epsilon}^{(1)} \geq \mu_{\textsc{E},\epsilon}^{(1)} \geq \mu_{\textsc{S}}^{(1)}$, and the optimal reward $\mu_*^{(1)}$ against a Bounded Memory player may be greater than $\mu_{\textsc{B},\epsilon}^{(1)}$ or in between these values, $\{\phi_j\}_{j}$ prioritizes the order of experts based on the optimal average reward they could achieve against the corresponding player 2 class (line \ref{initline}).

\section{Analysis}

We will now show that LAFF meets our key criteria of adaptability
and non-exploitability.
See Appendix for proofs of lemmas and the detailed proof of Theorem \ref{hedge}.
Lemma \ref{followregret} shows that
with high probability player 2's rewards against $\phi_E$ are not much greater than the EBS
(thus non-exploitability is feasible),
and player 1's rewards against a Follower are near the target when the correct Leader is used.

\begin{algorithm}
\caption{Lead and Follow Fairly (LAFF)}\label{followfirst}
\begin{algorithmic}[1]
\State \textbf{Init} target schedule $\{\mu_j\}_j = \{\mu_{\textsc{B},\epsilon}^{(1)}, \mu_{\textsc{B},\epsilon}^{(1)},\mu_{\textsc{E},\epsilon}^{(1)},\mu_{\textsc{E},\epsilon}^{(1)},$ $\mu_{\textsc{S}}^{(1)}\}$, expert schedule $\{\phi_j\}_j = \{\phi_F, \phi_B, \phi_F, \phi_E,$ $\phi_F, \phi_M\}$, expert index $j = 1$, $\tau = 0$, $R_\tau = 0$
\label{initline}
\For{$i=1,2,\dots,\ceil{T/H}$}
    \For{$t=(i-1)H + 1,\dots,\min\{iH, T\}$}
        \State Run expert $\phi_j$
        \State $R_\tau \leftarrow R_\tau + \mathbf{R}^{(1)}(A^{(1)}_t, A^{(2)}_t)$ \label{record}
    \EndFor
    \State $\tau \leftarrow \tau + H$ \label{tauup}
    \If{$j < |\{\phi_j\}_j|$ and $\frac{R_\tau}{\tau} < \mu_j - \mathcal{B}(\tau)$} \label{baselinecheck}
        \State $j \leftarrow j +1$, $\tau \leftarrow 0$, $R_\tau \leftarrow 0$
    \EndIf
\EndFor
\end{algorithmic}
\end{algorithm}

\begin{lemma}
\label{followregret}
\textbf{(Reward Bounds When LAFF Leads)} 
If player 1 uses $\phi_E$ over a sequence of length $\tau+K'$ starting at time $t^*+1$, then 
with probability at least $1- \frac{3\delta}{T}$:
\begin{align*}
    &\sum_{t=t^*+K'+1}^{t^* + K' + \tau} R^{(2)}_t \leq K' + 1 + \tau\mu_{\textsc{E},\epsilon}^{(2)} + 3\sqrt{\textstyle{\frac{1}{2}}\tau\log(\frac{T}{\delta})}.
\end{align*}
If player 2 is a Follower with $V^{(2)} = 0$, and player 1 uses $\phi_B$, then with probability at least $1- \frac{5\delta}{T}$, we have $\overline{r}^{(1)}_{i,\tau} \geq \mu_{\textsc{B},\epsilon}^{(1)} - \mathcal{B}(\tau)$.
If $V^{(2)} = \mu_{\textsc{E},\epsilon}^{(2)}$, and player 1 uses $\phi_E$, then with probability at least $1- \frac{5\delta}{T}$, we have $\overline{r}^{(1)}_{i,\tau} \geq \mu_{\textsc{E},\epsilon}^{(1)} - \mathcal{B}(\tau)$.
\end{lemma}

Lemma \ref{conditional_experts} guarantees that with high probability, LAFF follows or uses the maximin strategy against non-exploitative players, and punishes exploitative players.

\begin{lemma}
\label{conditional_experts}
\textbf{(False Positive and Negative Control of Exploitation Test)} Consider a sequence of $k$ epochs each of length $H$.
Let $m^*_{F}$ or $m^*_{M}$ be, respectively, the index of the \textit{subepoch} within this sequence at the start of which $\phi_F$ or $\phi_M$ switches to punishing with $\phi_E$, if at all (if not, let $m^*_{F}$ or $m^*_{M} = \infty$). Let $\slackexpl \geq \frac{2\mathcal{R}_{Q}(H/2, \delta/T)}{H} + \sqrt{\frac{2S^2A\log(c_0/\delta)}{c_1H}}$, where $c_0, c_1$ are defined as in Theorem 5.1 of \citet{MT05}, and $\slackmaximin \geq \sqrt{\frac{\log(T/\delta)}{2(H/2-K)}} + \sqrt{\frac{64e\log(N_q/\delta^2)}{(1-\lambda)(H/2-K)}}$, where $\lambda$ and $N_q$ are constants with respect to time defined in Lemma \ref{raolemma} (see Appendix).

Then, suppose player 2 is Bounded Memory, and $\phi_F$ is used. If $\mu_*^{(1)} < V^{(1)} - \slackexpl$, then with probability at least $1-\delta$, $m^*_{F} \leq \ceil{\frac{H^{1/2}}{2}}$. If $\mu_*^{(1)} \geq V^{(1)}$, then with probability at most $\frac{kH^{1/2}\delta}{T}$, $m^*_{F} < \infty$. If $\phi_M$ is used, and $\mu_{\textsc{M}}^{(2)} > \mu_{\textsc{E},\epsilon}^{(2)}$, then with probability at least $1-\delta$, $m^*_{M} \leq \ceil{\frac{H^{1/2}}{2}}$.

Suppose player 2 is Adversarial, with a sequence of action distributions $\{\pi^{(2)}_t\}$ such that, for any $M \geq H^{1/2} - K$ and $i$, $\frac{1}{M} \sum_{t=i+1}^{i+M} {\mathbf{v}^{(1)}_{\textsc{M}}}^\intercal \mathbf{R}^{(2)} \pi^{(2)}_t \leq \mu_{\textsc{E},\epsilon}^{(2)} - \slackmaximin$. Then, if $\phi_M$ is used, with probability at most $\frac{kH^{1/2}\delta}{T}$, $m^*_{M} < \infty$.
\end{lemma}

Our main result, Theorem \ref{hedge}, claims that 1)
against each of our target classes,
LAFF achieves a regret bound of the same order
as Optimistic Q-learning
in single-agent MDPs \citep{optQ},
and 2) LAFF satisfies non-exploitability.

\begin{theorem}
\label{hedge}
Let $\mathcal{C}$ be the set of player 2 algorithms that are any of the following:
\begin{itemize}
    \item Adversarial, with a sequence of action distributions $\{\pi^{(2)}_t\}$ such that $\frac{1}{M} \sum_{t=i+1}^{i+M} {\mathbf{v}^{(1)}_{\textsc{M}}}^\intercal \mathbf{R}^{(2)} \pi^{(2)}_t \leq \mu_{\textsc{E},\epsilon}^{(2)} - \slackmaximin$ for any $M \geq T^{1/4}$ and $i$,
    \item Follower, with $V^{(2)} \in \{0, \mu_{\textsc{E},\epsilon}^{(2)}\}$, or
    \item Bounded Memory, with 
    $\mu_*^{(1)} \geq V^{(1)}$.
\end{itemize}
Let $\slackmaximin$ and $\slackexpl$ satisfy the conditions of Lemma \ref{conditional_experts}.
Then, with probability at least $1-5\delta$, LAFF satisfies:
\begin{align*}
    \max_{\mathcal{C}} \mathcal{R}(T) &= \mathcal{O}(\mathcal{R}_{Q}(T, \delta/T)).
\end{align*}
Further, with probability at least $1-6\delta$, LAFF is 
$(V^{(1)},\slackexpl)$-non-exploitable 
when there exists an enforceable EBS.
\end{theorem}

If there is no enforceable EBS, $\mu_{\textsc{E},\epsilon}^{(2)} = \mu_{\textsc{S}}^{(2)}$ and so we cannot guarantee player 2 does worse than $\mu_{\textsc{E},\epsilon}^{(2)}$ in expectation.
The class of Adversarial players for which Theorem \ref{hedge} holds is technically restrictive. However, non-exploitability requires that for each strategy (expert) used by our algorithm that could be exploited, including Conditional Maximin, we exclude from our target class some subset of opponents. That is, we cannot guarantee low Adversarial regret against players who receive more than the EBS value against maximin, because such players may exploit us.

\begin{sketch}
For each opponent class, we need to show that with high probability LAFF does not lock in to
a suboptimal expert for that class. If LAFF locks in to an expert for which the corresponding target value $\mu_j$ is \textit{greater} than the opponent's benchmark $\mu(\mathfrak{B})$, this implies LAFF consistently receives rewards such that ``regret'' with respect to $\mu_j$ grows like $\mathcal{R}_Q$, by design of $\mathcal{B}(\tau)$. But since the benchmark is less than $\mu_j$, the true regret is also bounded as desired.

We therefore only need to consider the cases of $\mu_j \leq \mu(\mathfrak{B})$. First, we know that each expert achieves at most $\mathcal{R}_Q$ regret against its target opponent class, by, respectively: the definitions of $\mathcal{R}_Q$ (for non-exploitative Bounded Memory) and maximin (for Adversarial), and Lemma \ref{followregret} (for Followers). 
Lemma \ref{conditional_experts} ensures with high probability that $\phi_F$ and $\phi_M$ do not switch to $\phi_E$ when not exploited, so they inherit the desired regret bounds.

Then, we need only show that once LAFF reaches the expert
whose target class matches the opponent
(thus guaranteeing low regret using that expert), with high probability LAFF does not switch. 
But
if using the corresponding expert gives LAFF low regret with respect to $\mu(\mathfrak{B}) \geq \mu_j$, then its rewards are sufficiently high that the condition for switching experts (line \ref{baselinecheck} of Algorithm \ref{followfirst}) never holds. The first claim of the theorem follows.

\begin{figure*}[ht]
    \centering
    \begin{tabular}{ccc}
       \ \ Unconditional Follower (Q-Learning) & \ \ \ \ \ \ \ \ Conditional Follower (LAFF) & \ \ \ \ \ \ \ Bounded Memory (FTFT) \\
       \includegraphics[width=5.3cm]{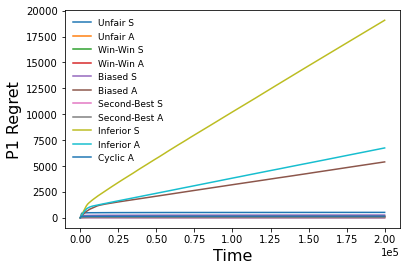} & \includegraphics[width=5.3cm]{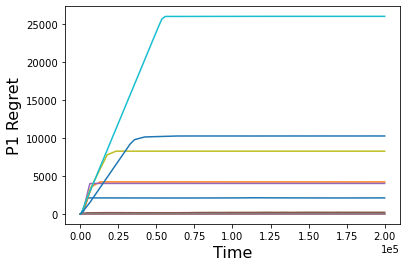} & \includegraphics[width=5.3cm]{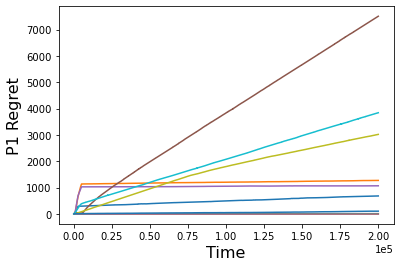} \\
    \end{tabular}
    \begin{tabular}{cc}
        \ \ \ \ \ \ \ \ \ Adversarial (Manipulator) & \ \  \ \ \ \ \ \  Exploitative (Bully) \\
        \includegraphics[width=5.3cm]{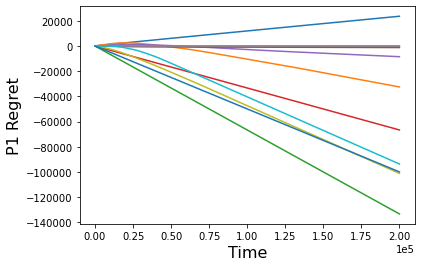} & \includegraphics[width=5.3cm]{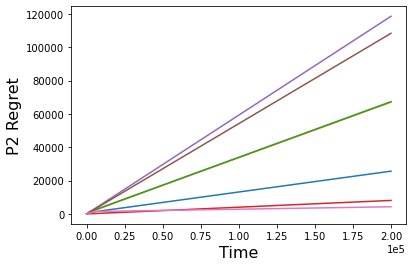} \\
    \end{tabular}
    \caption{The first four plots show LAFF's average regret, in each of 11 games detailed in the Appendix, for the following opponents: Unconditional Follower (Q-Learning), Conditional Follower (LAFF), Bounded Memory (FTFT), Adversarial (Manipulator). The last plot shows the regret of an Exploitative (Bully) algorithm against LAFF.}
    \label{fig:regrets}
\end{figure*}

To show non-exploitability, suppose LAFF locks in to the first instance of $\phi_F$. By Lemma \ref{conditional_experts}, $\phi_F$ detects evidence of exploitation sufficiently early that the remaining time left in the game is linear in $T$. After detecting exploitation, $\phi_F$ plays the same policy as $\phi_E$. But by Lemma \ref{followregret}, against this policy player 2 cannot guarantee an average reward greater than $\mu_{\textsc{E},\epsilon}^{(2)}$ plus a term that vanishes at a rate $T^{1/2}$. The second claim of the theorem follows for the other possible locked-in experts as well by considering two facts. First, whenever $\phi_E$ or $\phi_B$ is used, Lemma \ref{followregret} again bounds player 2's rewards, since by Pareto efficiency of the EBS player 2's rewards from the Bully solution cannot exceed $\mu_{\textsc{E},\epsilon}^{(2)}$. Second, if LAFF reaches $\phi_M$, again Lemma \ref{conditional_experts} ensures sufficiently fast detection of exploitation with high probability.
\end{sketch}

\section{Numerical Experiments}\label{sec:experiments}

Code for the experiments in this section is available on
Github.\footnote{\url{https://github.com/digiovannia/ad_expl}} 
We evaluate LAFF by three empirical metrics. First,
we find LAFF's empirical regret against one
algorithm from each target class.
Second, LAFF and a set of top-performing repeated games algorithms compete in a round-robin tournament. For each algorithm, we find its rewards against its best response algorithm in this set,
and check if it is in a learning equilibrium by applying a Nash equilibrium solver \citep{Knight2018} to the matrices of empirical rewards for algorithm pairs.
These criteria evaluate exploitability: more exploitable algorithms have lower rewards against algorithms that optimize against them,
and an exploitable algorithm cannot be in equilibrium with itself unless the fairness threshold $V^{(1)}$ is low. Finally, we perform a replicator dynamic simulation \citep{CO18}. Each generation, the algorithms' fitness values are computed as averages of the round-robin scores weighted by the distribution of the population of algorithms. Then, the population distribution is updated in proportion to fitness. This evaluates how well a given algorithm performs when the distribution of its opponents is determined by those algorithms' own performance. 
Exploitability is thus implicitly penalized by accounting for opponents' incentives.
Details on the implementation of these experiments are in the Appendix. We set $V^{(1)} = \mu_{\textsc{E},\epsilon}^{(1)}$.

Our set of competitors to LAFF consists of Bounded Memory (Bully, Forgiving Generalized Tit-for-Tat or FTFT), Follower (M-Qubed, Q-Learning, Fictitious Play), and expert (Manipulator, S++) algorithms. See Appendix for details and sources.
We chose these algorithms because, first, they performed
well
in a repeated games tournament \citep{CO18},
and second,
they cover our opponent classes.
S++ and Manipulator do not fall cleanly into any of those classes, but they are the closest comparisons in previous literature to LAFF, since they
adapt to a variety of opponents by switching between Leader and Follower experts.

To ensure sufficient diversity of test games, we choose games based on the taxonomy of Figure 1 in \citet{topology}. Six game families
are categorized by the structures of their Nash equilibria.
We use two games from each family, one with symmetric rewards and one with asymmetric, except Cyclic, which has no symmetric games (see Appendix). 

\begin{table*}
\centering
\caption{Rewards of algorithm pairs, averaged over games and trials (pure learning equilibria in are highlighted in bold text, and each algorithm's reward against its best response is in blue)}
\label{tab:emp_matrix}
\begin{tabular}{ccccccccc}
    \toprule
    & S++ & Manipulator & M-Qubed & Bully & Q-Learning & LAFF & FTFT & FP \\
    \midrule
    S++ & 0.75, 0.76 & 0.73, 0.80 & \textcolor{blue}{0.73}, 0.81 & 0.65, 0.77 & 0.82, 0.76 & 0.71, 0.8 & 0.70, 0.68 & 0.72, 0.55 \\
    Manipulator & 0.87, 0.68 & 0.76, 0.71 & 0.77, 0.65 & \textcolor{blue}{0.65}, 0.77 & 0.89, 0.67 & 0.70, 0.65 & 0.71, 0.60 & 0.76, 0.55 \\
    M-Qubed & 0.88, \textcolor{blue}{0.68} & 0.68, 0.68 & 0.80, 0.74 & \textcolor{blue}{0.65}, 0.80 & 0.79, 0.75 & 0.76, 0.73 & 0.78, 0.65 & 0.62, 0.56 \\
    Bully & 0.86, 0.61 & 0.83, \textcolor{blue}{0.60} & 0.85, \textcolor{blue}{0.61} & 0.48, 0.44 & \textbf{\textcolor{blue}{0.91}, \textcolor{blue}{0.63}} & 0.61, 0.49 & 0.72, 0.55 & 0.76, \textcolor{blue}{0.56} \\
    Q-Learning & 0.82, 0.77 & 0.73, 0.83 & 0.79,  0.67& \textbf{\textcolor{blue}{0.68}, \textcolor{blue}{0.85}} & 0.83, 0.74 & 0.71, 0.84 & 0.81, \textcolor{blue}{0.67} & 0.64, 0.56 \\
    LAFF & 0.87, 0.65 & 0.71, 0.66 & 0.74, 0.72 & 0.55, 0.61 & 0.90, 0.66 & \textbf{\textcolor{blue}{0.77}, \textcolor{blue}{0.74}} & 0.80, 0.70 & 0.75, 0.57 \\
    FTFT & 0.64, 0.70 & 0.49, 0.71 & 0.59, 0.76& 0.60, 0.71 & 0.59, 0.78 & \textcolor{blue}{0.61}, 0.78 & 0.80, 0.75 & 0.46, 0.72 \\
    FP & 0.70, 0.73 & \textcolor{blue}{0.66}, 0.74 & 0.66, 0.55 & 0.63, 0.73 & 0.69, 0.57 & 0.61, 0.71 & 0.71, 0.60 & 0.68, 0.55 \\
    \bottomrule
\end{tabular}
\end{table*}


\paragraph{\textbf{Regret Bounds}} Figure \ref{fig:regrets} shows LAFF's regret, averaged over 50 trials, in games against an algorithm from each target class, and the regret of an exploitative Bounded Memory algorithm against LAFF.
We chose Manipulator as ``Adversarial'' because it does not play the EBS and is not a pure Leader or Follower.
However, in the symmetric Unfair game, the empirical rewards indicate that Manipulator attempts to exploit LAFF,
so LAFF punishes Manipulator at the expense of the Adversarial regret guarantee.
From the plot evaluating player 2's regret, we also exclude four games where player 2's Bully solution equals the EBS, since in these cases $\mu_*^{(1)} \geq V^{(1)}$
(player 1 is not exploited by playing the optimal policy).
In most games, LAFF's regret
eventually plateaus,
while the exploitative player has linear regret, showing that LAFF is non-exploitable.
In three games, 
LAFF has linear regret against an Unconditional Follower and non-exploitative Bounded Memory player. This may be due to the practical difficulty of choosing hyperparameters for tests used to decide when to switch to the next expert; these tests depend on some unknown quantities, so for our experiments, we tuned $\mathcal{B}(\tau)$ on a training set of four games that are not included in the set of 11 games for these results
(see Appendix).
Longer time horizons may be required for
the conditions on $\slackexpl$ in Lemma \ref{conditional_experts} to hold.
We used a horizon of $T=2 \cdot 10^5$ to be on
the same approximate scale as experiments in other works on repeated games \citep{CG10, LS05, C14}.

\begin{figure}[ht]
    \centering
    \includegraphics[width=7cm]{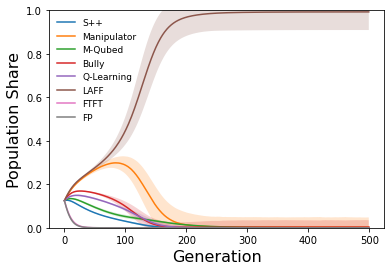}
    \caption{Replicator dynamic results, where the bold curves are average population shares and shaded regions are plus and minus one standard deviation.}
    \label{fig:rd_results}
\end{figure}

\paragraph{\textbf{Round Robin}} Table \ref{tab:emp_matrix} shows the average rewards of each algorithm pair across the 11 games and 50 trials,
which provide an empirical bimatrix for the \textit{learning game}, i.e., a meta-game in which users choose algorithms to deploy across different repeated games.
An algorithm's reward against its best response (highlighted in blue) measures how much it bullies when possible and avoids exploitation.
Both as player 1 and player 2, LAFF is second by this metric, behind Bully. We also highlight the pure strategy Nash equilibria of this learning game (in bold), noting that LAFF is in a learning equilibrium with itself. Unfortunately, the pairing in which Q-Learning follows Bully is also an equilibrium. Thus there is an equilibrium selection problem, e.g., both users might choose Bully and receive very low rewards. However, in practice it may be easier for users to coordinate on both using LAFF, because there is no conflict over choosing which side is the Leader (Bully) versus the Follower (Q-Learning).


\paragraph{\textbf{Replicator Dynamic}}
On average over 1000 runs,
LAFF converges to 100\% of the population in the pool of algorithms (Figure \ref{fig:rd_results}), based on fitness computed as the \textit{minimum} of an algorithm's average reward over the set of games when playing as player 1 versus player 2. This metric matches the motivation
for the EBS; algorithm users will not know \textit{a priori} which of the two ``sides'' of the game they will be in. Thus, they may prefer their algorithm to cooperate with itself (maximize an egalitarian objective), instead of bullying its copy in hopes of being on the side of the bully.


\section{Discussion}

When choosing algorithms for multi-agent interactions, users
will have to trade off robustness to the variety
of possible algorithms they might face, with avoiding providing other users incentives to exploit them \citep{SLRC21}.
We have presented an algorithm for repeated games that balances these desiderata.
Both properties can facilitate cooperation between learning agents, while still allowing them to accept generous offers.
If LAFF faces an agent who ``follows'' fair, Pareto efficient bargaining proposals, the Egalitarian Leader leads them to a mutual benefit over their security values.
If the other agent's fairness standard is different, the Conditional Follower can follow this alternative proposal using RL if it is not exploitative;
otherwise, the exploitation penalty encourages the other player to be more cooperative.
Against exploitable agents, the Bully Leader can benefit from a more self-interested bargain.
Finally, if the other player is unwilling to cooperate at all but is not exploitative, Conditional Maximin ensures safety. In future work, more experts can be added
based on agent classes that we have neglected. For example, while LAFF includes Leader experts only for the extreme cases in which player 2 has a high or minimal fairness standard, one could add Leaders for other bargaining solutions.

The biggest limitations of our approach are restrictive assumptions required for our non-exploitability criterion, and the strictness of this criterion. The margin $\slackexpl$ is small only for sufficiently large time horizons,
hence the linear regret in some of our experiments. Though LAFF successfully punishes players against whom it receives less than fair rewards, this is only strategically necessary when such players \textit{benefit} from playing this way (genuine ``exploitation'').
It may not be practically necessary
to modify the experts to not punish
when the opponent also does worse,
because an opponent would not have an incentive to lead with a Pareto inefficient policy.
Finally, we note that 
our approach is not intended to provide the optimal balance of the adaptability-exploitability tradeoff;
in particular, keeping a fixed fairness threshold
may not be ideal if it
prevents an algorithm from
cooperating with algorithms
that follow other intuitively ``fair'' standards \citep{SLRC21}.

\begin{contributions} 
    Both authors conceived and carried out the research project jointly. A.D.~wrote the paper and code for
    numerical experiments. A.T.~helped edit the paper.
\end{contributions}

\begin{acknowledgements} 
    A.D.~acknowledges the support of a grant
    from the Center on Long-Term Risk Fund.
\end{acknowledgements}

\bibliography{digiovanni_111}

\begin{thebibliography}{26}
\providecommand{\natexlab}[1]{#1}
\providecommand{\url}[1]{\texttt{#1}}
\expandafter\ifx\csname urlstyle\endcsname\relax
  \providecommand{\doi}[1]{doi: #1}\else
  \providecommand{\doi}{doi: \begingroup \urlstyle{rm}\Url}\fi

\bibitem[Brafman and Tennenholtz(2004)]{BT04}
Ronen~I. Brafman and Moshe Tennenholtz.
\newblock Efficient learning equilibrium.
\newblock \emph{Artificial Intelligence}, 159:\penalty0 27--47, 2004.

\bibitem[Brown(1951)]{fictplay}
George~W. Brown.
\newblock Iterative solutions of games by fictitious play.
\newblock In T.~C. Koopmans, editor, \emph{Activity Analysis of Production and
  Allocation}. Wiley, 1951.

\bibitem[Bruns(2010)]{topology}
Bryan Bruns.
\newblock Navigating the topology of 2x2 games: an introductory note on payoff
  families, normalization, and natural order, 2010.
\newblock URL \url{https://arxiv.org/abs/1010.4727}.

\bibitem[Chakraborty and Stone(2010)]{ICML10-chakraborty}
Doran Chakraborty and Peter Stone.
\newblock Convergence, targeted optimality and safety in multiagent learning.
\newblock \emph{Proceedings of the Twenty-seventh International Conference on
  Machine Learning (ICML)}, 2010.

\bibitem[Clifton and Riché(2021)]{CR21}
Jesse Clifton and Maxime Riché.
\newblock Towards cooperation in learning games, 2021.

\bibitem[Crandall(2014)]{C14}
Jacob~W. Crandall.
\newblock Towards minimizing disappointment in repeated games.
\newblock \emph{Journal of Artificial Intelligence Research}, 49:\penalty0
  111--142, 2014.

\bibitem[Crandall(2020)]{C20}
Jacob~W. Crandall.
\newblock When autonomous agents model other agents: An appeal for altered
  judgment coupled with mouths, ears, and a little more tape.
\newblock \emph{Artificial Intelligence}, 2020.

\bibitem[Crandall and Goodrich(2010)]{CG10}
Jacob~W. Crandall and Michael~A. Goodrich.
\newblock Learning to compete, coordinate, and cooperate in repeated games
  using reinforcement learning.
\newblock \emph{Machine Learning}, 82:\penalty0 281--314, 2010.

\bibitem[Crandall et~al.(2018)Crandall, Oudah, Tennom, Ishowo-Oloko, Abdallah,
  Bonnefon, Cebrian, Shariff, Goodrich, and Rahwan]{CO18}
Jacob~W. Crandall, Mayada Oudah, Tennom, Fatimah Ishowo-Oloko, Sherief
  Abdallah, Jean-François Bonnefon, Manuel Cebrian, Azim Shariff, Michael~A.
  Goodrich, and Iyad Rahwan.
\newblock Cooperating with machines.
\newblock \emph{Nature Communications}, 9:\penalty0 233, 2018.

\bibitem[Jacq et~al.(2020)Jacq, Perolat, Geist, and Pietquin]{fcl}
Alexis Jacq, Julien Perolat, Matthieu Geist, and Olivier Pietquin.
\newblock Foolproof cooperative learning.
\newblock In Sinno~Jialin Pan and Masashi Sugiyama, editors, \emph{Proceedings
  of The 12th Asian Conference on Machine Learning}, volume 129 of
  \emph{Proceedings of Machine Learning Research}, pages 401--416, Bangkok,
  Thailand, 18--20 Nov 2020. PMLR.
\newblock URL \url{http://proceedings.mlr.press/v129/jacq20a.html}.

\bibitem[Knight and Campbell(2018)]{Knight2018}
Vincent Knight and James Campbell.
\newblock Nashpy: A {Python} library for the computation of {Nash} equilibria.
\newblock \emph{Journal of Open Source Software}, 3\penalty0 (30):\penalty0
  904, 2018.
\newblock \doi{10.21105/joss.00904}.
\newblock URL \url{https://doi.org/10.21105/joss.00904}.

\bibitem[Littman and Stone(2001)]{LS01}
Michael Littman and Peter Stone.
\newblock Implicit negotiation in repeated games.
\newblock \emph{Revised Papers from the 8th International Workshop on
  Intelligent Agents VIII}, 2001.

\bibitem[Littman and Stone(2005)]{LS05}
Michael~L. Littman and Peter Stone.
\newblock A polynomial-time {Nash} equilibrium algorithm for repeated games.
\newblock \emph{Decision Support Systems}, 39:\penalty0 55--66, 2005.

\bibitem[Mailath and Samuelson(2006)]{MS06}
George Mailath and Larry Samuelson.
\newblock \emph{Repeated Games and Reputations: Long-Run Relationships}.
\newblock Oxford University Press, 2006.

\bibitem[Mannor and Tsitsiklis(2005)]{MT05}
Shie Mannor and John~N. Tsitsiklis.
\newblock On the empirical state-action frequencies in {Markov} decision
  processes under general policies.
\newblock \emph{Mathematics of Operations Research}, 30:\penalty0 545--561,
  2005.

\bibitem[Paulin(2015)]{P18}
Daniel Paulin.
\newblock Concentration inequalities for {Markov} chains by {Marton} couplings
  and spectral methods.
\newblock \emph{Electronic Journal of Probability}, 20:\penalty0 1--32, 2015.

\bibitem[Powers and Shoham(2004)]{PS04}
Rob Powers and Yoav Shoham.
\newblock New criteria and a new algorithm for learning in multi-agent systems.
\newblock \emph{Neural Information Processing Systems}, 2004.

\bibitem[Powers and Shoham(2005)]{PS05}
Rob Powers and Yoav Shoham.
\newblock Learning against opponents with bounded memory.
\newblock \emph{Proceedings of the 19th International Joint Conference on
  Artificial Intelligence}, 2005.

\bibitem[Press and Dyson(2012)]{Press10409}
William~H. Press and Freeman~J. Dyson.
\newblock Iterated prisoner{\textquoteright}s dilemma contains strategies that
  dominate any evolutionary opponent.
\newblock \emph{Proceedings of the National Academy of Sciences}, 109\penalty0
  (26):\penalty0 10409--10413, 2012.
\newblock ISSN 0027-8424.
\newblock \doi{10.1073/pnas.1206569109}.
\newblock URL \url{https://www.pnas.org/content/109/26/10409}.

\bibitem[Puterman(1994)]{puterman}
Martin~L. Puterman.
\newblock \emph{Markov Decision Processes: Discrete Stochastic Dynamic
  Programming}.
\newblock John Wiley \& Sons, Inc., New York, NY, USA, 1994.

\bibitem[Rao(2019)]{Rao19}
Shravas Rao.
\newblock {A Hoeffding inequality for Markov chains}.
\newblock \emph{Electronic Communications in Probability}, 24:\penalty0 1--11,
  2019.

\bibitem[Stastny et~al.(2021)Stastny, Riché, Lyzhov, Treutlein, Dafoe, and
  Clifton]{SLRC21}
Julian Stastny, Maxime Riché, Alexander Lyzhov, Johannes Treutlein, Allan
  Dafoe, and Jesse Clifton.
\newblock Normative disagreement as a challenge for cooperative {AI}, 2021.

\bibitem[Stewart and Plotkin(2012)]{SPtournament}
Alexander~J. Stewart and Joshua~B. Plotkin.
\newblock Extortion and cooperation in the prisoner’s dilemma.
\newblock \emph{Proceedings of the National Academy of Sciences}, 109, 2012.

\bibitem[Tossou et~al.(2020)Tossou, Dimitrakakis, Rzepecki, and Hofmann]{TD20}
Aristide~C.Y. Tossou, Christos Dimitrakakis, Jaroslaw Rzepecki, and Katja
  Hofmann.
\newblock A novel individually rational objective in multi-agent multi-armed
  bandits: Algorithms and regret bounds.
\newblock \emph{Proc. of the 19th International Conference on Autonomous Agents
  and Multiagent Systems}, 2020.

\bibitem[Watkins and Dayan(1992)]{WD92}
Chris~J. Watkins and Peter Dayan.
\newblock Q-learning.
\newblock \emph{Machine Learning}, 8:\penalty0 279--292, 1992.

\bibitem[Wei et~al.(2020)Wei, Jafarnia-Jahromi, Luo, Sharma, and Jain]{optQ}
Chen-Yu Wei, Mehdi Jafarnia-Jahromi, Haipeng Luo, Hiteshi Sharma, and Rahul
  Jain.
\newblock Model-free reinforcement learning in infinite-horizon average-reward
  {Markov} decision processes.
\newblock volume 119 of \emph{Proceedings of the 37th International Conference
  on Machine Learning}, 2020.

\end{thebibliography}

\clearpage

\title{Balancing Adaptability and Non-exploitability in Repeated Games \\
(Supplementary Material)}

\onecolumn
\maketitle

\appendix


\section{Details on the Formal Setting}\label{formal}

The randomization weights $w_t^{(i)}$ introduced in Section \ref{sec:sub:rgclass} technically induce an infinite action space for both players. However, as discussed in Section \ref{sec:sub:regretdef}, the benchmarks in our problem statement are not defined with respect to the globally optimal policy in a repeated game, if such an object is even well-defined. Against a Bounded Memory player, who commits to a fixed $w_t^{(2)}$, the optimal policy is equivalent to that for an MDP, and is independent of $w_t^{(1)}$. Against an Adversarial player, the maximin strategy also does not depend on $w_t^{(i)}$. Finally, against Followers, benchmarks are defined with respect to bargaining solutions that only require a constant $w_t^{(1)}$. Therefore, it is not necessary for our purposes to consider the infinite action space of possible $w_t^{(1)}$ values that player 1 can choose at each time step.

\section{Derivation of Enforceable EBS and Bully Solution}
\label{app:derivation}

 While it has been shown that the EBS can be tractably computed absent enforceability constraints \citep{TD20}, it is nontrivial that this extends to the constrained case.
Lemma \ref{enforce}
helps us construct the enforceability-constrained EBS.

\begin{lemma}
\label{enforce}
Consider any function $f$ that is monotone in $\mathcal{U}$, that is, if $u_1 \geq v_1$ and $u_2 \geq v_2$ then $f(u_1,u_2) \geq f(v_1,v_2)$. Then there always exists a maximizer of $f$ over $\mathcal{U}(\epsilon)$ that is a convex combination of no more than two points in $\mathcal{G}$.
\end{lemma}

\begin{proof}
The argument is similar to that in \citet{LS05}. Let $(u_1, u_2)$ be any point in $\mathcal{U}(\epsilon)$, and suppose that any point $(u_1',u_2')$ with $u_1' \geq u_1$ and $u_2' \geq u_2$ (except $(u_1, u_2)$ itself) is not in $\mathcal{U}(\epsilon)$. Then either $(u_1,u_2)$ is on the boundary of $\mathcal{U}$, or it is in the interior and all points to its upper-right quadrant (denoted $Q_{u_1,u_2}$) are excluded by enforceability. By convexity, the former implies $(u_1, u_2)$ is a convex combination of no more than two points in $\mathcal{G}$. If the latter, $Q_{u_1,u_2} \cap \mathcal{U}$ must be a subset of the whole region excluded by enforceability for some set of points $\mathcal{X}$, that is, $\text{Conv}(\mathcal{X}) \cap \{(-\infty, \infty) \times (-\infty, v(\epsilon, K, \mathcal{X}))\}$ for some $v(\epsilon, K, \mathcal{X})$. But this again implies the desired conclusion, because $(u_1,u_2)$ must be on a boundary of that excluded region other than the one induced by $\{(-\infty, \infty) \times (-\infty, v(\epsilon, K, \mathcal{X}))\}$.
\end{proof}

The $\epsilon$-enforceable EBS, which we will use to design one of the Leader experts, is found as follows. Assign to each joint action pair $x_A := (i_1, j_1)$ and $x_B := (i_2, j_2)$ the score $\rho(x_A,x_B) := \max_{\alpha_{AB}} \min_{i=1,2}\{\alpha_{AB} \mathbf{R}^{(i)}(x_A) + (1-\alpha_{AB})\mathbf{R}^{(i)}(x_B) - \mu_{\textsc{S}}^{(i)}\}$, where $\mathbf{R}^{(i)}(x_A) := \mathbf{R}^{(i)}(i_1,j_1)$ and $\mathbf{R}^{(i)}(x_B) := \mathbf{R}^{(i)}(i_2,j_2)$, and choose the pair with the highest score \citep{TD20}.
Searching over pairs is sufficient by Lemma \ref{enforce}. We maximize $\rho$ over $\alpha_{AB}$ subject to enforceability. 
For two points such that $\mathbf{R}^{(2)}(x_A) > \mathbf{R}^{(2)}(x_B)$ (order does not matter), $\epsilon$-enforceability requires:
\begin{align*}
    & \alpha_{AB} \geq \frac{ r(\{x_A, x_B\}) + \epsilon + K[\mu_{\textsc{S}}^{(2)} - \mathbf{R}^{(2)}(x_B)]}{K [\mathbf{R}^{(2)}(x_A) - \mathbf{R}^{(2)}(x_B)]}.
\end{align*}
If $\mathbf{R}^{(2)}(x_A) = \mathbf{R}^{(2)}(x_B)$, then $\alpha_{AB}$ can be arbitrary as long as the first line above still holds; otherwise, this pair is not enforceable regardless of $\alpha_{AB}$.
Taking $\mathbf{R}^{(2)}(x_A) > \mathbf{R}^{(2)}(x_B)$ without loss of generality,
there are two cases to consider.
(1) If $\mathbf{R}^{(i)}(x_A) \geq \mathbf{R}^{(i)}(x_B)$ for both $i=1,2$,
both functions in the minimum have nonnegative slope, so $\rho$ is nondecreasing in $\alpha_{AB}$.
Otherwise, (2) $\rho$ has its maximum at $a = \frac{\mathbf{R}^{(2)}(x_B) - \mathbf{R}^{(1)}(x_B)}{\mathbf{R}^{(1)}(x_A) - \mathbf{R}^{(1)}(x_B) + \mathbf{R}^{(2)}(x_B) - \mathbf{R}^{(2)}(x_A)}$.

In case 1, since $\epsilon$-enforceability is a \textit{lower} bound $v(\epsilon, K)$ on $\alpha_{AB}$, the optimal $\alpha_{AB} = 1$ if that upper bound is at most 1, otherwise this pair is not enforceable.
In case 2, if enforceability does not exclude $a$, then $\alpha_{AB} = a$. Otherwise, the non-excluded region must decrease down from $v(\epsilon, K)$ or increase up to $v(\epsilon, K)$; either way, $\alpha_{AB} = v(\epsilon, K)$ is optimal.

Finally, we also construct the Bully solution for the second Leader expert by following the procedure above, except with a ``selfish'' score $\rho(x_A,x_B) := \max_{\alpha_{AB}} \alpha_{AB} \mathbf{R}^{(1)}(x_A) + (1-\alpha_{AB})\mathbf{R}^{(1)}(x_A)$.
This is, again, a monotone function over $\mathcal{U}(\epsilon)$, so searching over pairs of joint actions suffices. If $\mathbf{R}^{(1)}(x_A) \leq \mathbf{R}^{(1)}(x_B)$, $\rho$ is nondecreasing in $\alpha_{AB}$, so as before we set
$\alpha_{AB} = v(\epsilon, K)$.
If $\mathbf{R}^{(1)}(x_A) > \mathbf{R}^{(1)}(x_B)$, we set $\alpha_{AB} = 1$.

\section{Proof of Lemma \ref{followregret}}

\begin{replemma}{followregret}
\textbf{(Reward Bounds When LAFF Leads)} Let $t^*+1$ be the start time of a sequence of time steps of total length $\tau+K'$. If player 1 uses $\phi_E$ over this sequence, then with probability at least $1- \frac{3\delta}{T}$:
\begin{align*}
    &\sum_{t=t^*+K'+1}^{t^* + K' + \tau} R^{(2)}_t \leq K' + 1 + \tau\mu_{\textsc{E},\epsilon}^{(2)} + 3\sqrt{\textstyle{\frac{1}{2}}\tau\log(\frac{T}{\delta})}.
\end{align*}
Further, if player 2 is a Follower with $V^{(2)} = 0$, and player 1 uses $\phi_B$, then with probability at least $1- \frac{5\delta}{T}$, we have $\overline{r}^{(1)}_{i,\tau} \geq \mu_{\textsc{B},\epsilon}^{(1)} - \mathcal{B}(\tau)$.
If $V^{(2)} = \mu_{\textsc{E},\epsilon}^{(2)}$, and player 1 uses $\phi_E$, then with probability at least $1- \frac{5\delta}{T}$, we have $\overline{r}^{(1)}_{i,\tau} \geq \mu_{\textsc{E},\epsilon}^{(1)} - \mathcal{B}(\tau)$.
\end{replemma}

\begin{proof}
We define the function $\mathcal{B}$ that controls the false positive rate of LAFF's hypothesis tests as follows:
\begin{align*}
    \xi(\epsilon, r) &:= \begin{cases}
    \frac{\epsilon}{2K'},& \text{if } r \geq 0\\
    \frac{\epsilon + r}{2K'},& \text{if } -\epsilon < r < 0\\
    -r,              & \text{otherwise},
\end{cases} \\
    \mathcal{B}(\tau) &:= \frac{1}{\tau} \cdot \frac{K'\xi(\epsilon, r(\mathcal{X})) + C_1T_0 + K'+1}{\xi(\epsilon, r(\mathcal{X}))} \\
    &+ \frac{1}{\tau} \cdot \frac{C_2\mathcal{R}_{Q}(\tau, \frac{\delta}{T}) + (3 + \xi(\epsilon, r(\mathcal{X})))\sqrt{\frac{\tau \log(\frac{T}{\delta})}{2}}}{\xi(\epsilon, r(\mathcal{X}))}.
\end{align*}
Where $\mathcal{X} = \mathcal{X}_{\textsc{B}} := \{(a_{\textsc{B}}^{(1)}(y), a_{\textsc{B}}^{(2)}(y))\}_{y=0,1}$ for expert index $j \leq 2$, $\mathcal{X} = \mathcal{X}_{\textsc{E}} := \{(a_{\textsc{E}}^{(1)}(y), a_{\textsc{E}}^{(2)}(y))\}_{y=0,1}$ for $j > 2$, and $\delta > 0$ is some confidence level.

First suppose $V^{(2)} = \mu_{\textsc{E},\epsilon}^{(2)}$. Consider the target action pair $\mathcal{X}_{\textsc{E}}$ and weight $\alpha_{\textsc{E}}$. Note that after the first ${K'}$ time steps, $\phi_E$ is stationary and thus induces a communicating MDP from player 2's perspective, with optimal average reward $\mu_*^{(2)}$. We have that $\mu_*^{(2)} \geq \mu_{\textsc{E},\epsilon}^{(2)}$ when player 2 plays against $\phi_E$. To see this, note that the policy of playing $a_{\textsc{E}}^{(2)}(y_t^{(1)})$ for all times $t$, when player 1 uses $\phi_E$, induces a Markov reward process defined by two ``states'' $\{0, 1\}$, with rewards $(\mathbf{R}^{(2)}(a_{\textsc{E}}^{(1)}(0), a_{\textsc{E}}^{(2)}(0)), \mathbf{R}^{(2)}(a_{\textsc{E}}^{(1)}(1), a_{\textsc{E}}^{(2)}(1)))$ and the following transition matrix:
\begin{align*}
    \left[\begin{tabular}{cc}
        $\alpha_{\textsc{E}}$ & $1-\alpha_{\textsc{E}}$ \\
        $\alpha_{\textsc{E}}$ & $1-\alpha_{\textsc{E}}$
    \end{tabular}\right].
\end{align*}
This process has stationary distribution $(\alpha_{\textsc{E}}, 1-\alpha_{\textsc{E}})$, hence the limit average reward of this policy is $\mu_{\textsc{E},\epsilon}^{(2)}$.

For time $t$, let $D_t$ be the event that $\phi_E$ is not following $\mathbf{v}^{(1)}_P$ and $A^{(2)}_{t} \neq a_{\textsc{E}}^{(2)}(Y_{t}^{(1)})$, and $M_t$ be the event that $\phi_E$ is following $\mathbf{v}^{(1)}_P$. Respectively, these represent the events that $\phi_E$ is not punishing but player 2 deviates from the target solution, and that $\phi_E$ is punishing.  Define the random set $\mathcal{T} := \{t \in \{t_i+K',\dots,t_i + \tau - {K'}\} \ | \ D_t\}$, and for each $t \in \mathcal{T}$, let $\tau_t$ be the first time $t'>t$ such that $M_{t'}$ holds but $M_{t'+1}$ does not.

For simplicity of notation, reindex $t_i = 0$, and let $\tau_+ = \sum_{t=1}^\tau \mathbb{I}[D_t^c \cap M_t^c]$ and $\tau_{+,{K'}} = \sum_{t={K'}+1}^{\tau} \mathbb{I}[D_t^c \cap M_t^c]$. Define $r_{j}^{(i)} := \mathbf{R}^{(i)}(a_{\textsc{E}}^{(1)}(j), a_{\textsc{E}}^{(2)}(j))$ for players $i=1,2$ and targets $j=0,1$.
WLOG, let $r_{0}^{(i)} \geq r_{1}^{(i)}$. Conditional on $D_t^c \cap M_t^c$, we have $R^{(i)}_t \stackrel{iid}{\sim} r_{1}^{(i)} + (r_{0}^{(i)} - r_{1}^{(i)})\text{Bern}(\alpha_{\textsc{E}})$, and $\mu_{\textsc{E},\epsilon}^{(i)} = r_{1}^{(i)} + (r_{0}^{(i)} - r_{1}^{(i)})\alpha_{\textsc{E}}$. Then, by Hoeffding's inequality:
\begin{align*}
    P\left( \sum_{t=1}^{\tau} R^{(1)}_t\mathbb{I}[D_t^c \cap M_t^c] \leq \tau_+ \left(\mu_{\textsc{E},\epsilon}^{(1)} - (r_{0}^{(1)} - r_{1}^{(1)})\sqrt{\frac{\log(T/\delta)}{2\tau_+}}\right)\right) \leq \frac{\delta}{T}, \\
    P\left( \sum_{t={K'}+1}^{\tau} R^{(2)}_t\mathbb{I}[D_t^c \cap M_t^c] \geq \tau_{+,{K'}} \left(\mu_{\textsc{E},\epsilon}^{(2)} + (r_{0}^{(2)} - r_{1}^{(2)})\sqrt{\frac{\log(T/\delta)}{2\tau_{+,K'}}}\right)\right) \leq \frac{\delta}{T}.
\end{align*}
Since $\phi_E$ only conditions on the past after the first $K'$ time steps, there are no punishments for actions player 2 may have taken prior to time $t=1$. This guarantees that, for $t \geq K'$, any event $M_t$ must be preceded by either $M_{t-1}$ or $D_{t-1}$ for $t-1 \geq K'$.
Therefore $\sum_{t={K'}+1}^{\tau} R_t^{(2)}\mathbb{I}[D_t \cup M_t] \leq K' + \sum_{t\in \mathcal{T}} (R_t^{(2)} + \sum_{t'=t+1}^{\tau_t} R_{t'}^{(2)})$. So with probability at least $1-\frac{2\delta}{T}$:
\begin{align*}
    &\tau(\mu_{\textsc{E},\epsilon}^{(1)} - \overline{r}^{(1)}_{i,\tau}) = \sum_{t=1}^{\tau} (\mu_{\textsc{E},\epsilon}^{(1)} - R^{(1)}_t)\mathbb{I}[D_t \cup M_t] + \sum_{t=1}^{\tau}  (\mu_{\textsc{E},\epsilon}^{(1)} - R^{(1)}_t)\mathbb{I}[D_t^c \cap M_t^c] \\
    &\qquad \leq \sum_{t=1}^{\tau} (\mu_{\textsc{E},\epsilon}^{(1)} - R^{(1)}_t)\mathbb{I}[D_t \cup M_t] + \tau_+\mu_{\textsc{E},\epsilon}^{(1)} - \tau_+ \left[r_{1}^{(1)} + (r_{0}^{(1)} - r_{1}^{(1)})\left(\alpha_{\textsc{E}} - \sqrt{\frac{\log(T/\delta)}{2\tau_+}}\right)\right] \\
    &\qquad \leq \sum_{t=1}^{\tau} \mathbb{I}[D_t \cup M_t] + \tau_+ (r_{0}^{(1)} - r_{1}^{(1)})\sqrt{\frac{\log(T/\delta)}{2\tau_+}}, \\
 &(\tau-{K'})(\mu_{\textsc{E},\epsilon}^{(2)} - \overline{r}^{(2)}_{i,\tau}) \\
   &\qquad \geq \sum_{t={K'}+1}^{\tau}(\mu_{\textsc{E},\epsilon}^{(2)} - R_t^{(2)})\mathbb{I}[D_t \cup M_t] + \tau_{+,{K'}}\mu_{\textsc{E},\epsilon}^{(2)} - \tau_{+,{K'}} \left[r_{1}^{(2)} + (r_{0}^{(2)} - r_{1}^{(2)})\left(\alpha_{\textsc{E}} + \sqrt{\frac{\log(T/\delta)}{2\tau_{+,{K'}}}}\right)\right] \\
   &\qquad = \sum_{t={K'}+1}^{\tau}(\mu_{\textsc{E},\epsilon}^{(2)} - R_t^{(2)})\mathbb{I}[D_t \cup M_t] - \tau_{+,{K'}}(r_{0}^{(2)} - r_{1}^{(2)})\sqrt{\frac{\log(T/\delta)}{2\tau_{+,{K'}}}}\\
   &\qquad \geq \mu_{\textsc{E},\epsilon}^{(2)}\sum_{t={K'}+1}^{\tau}\mathbb{I}[D_t \cup M_t] - {K'} - \sum_{t\in \mathcal{T}} \left(R_t^{(2)} + \sum_{t'=t+1}^{\tau_t} R_{t'}^{(2)}\right) - \tau_{+,{K'}}(r_{0}^{(2)} - r_{1}^{(2)})\sqrt{\frac{\log(T/\delta)}{2\tau_{+,{K'}}}}.
\end{align*}
Let $R_{\textsc{E},t}^{(2)} := \mathbf{R}^{(2)}(a_{\textsc{E}}^{(1)}(Y_{t}^{(1)}), a_{\textsc{E}}^{(2)}(Y_{t}^{(1)}))$. By enforceability, $r(\mathcal{X}_{\textsc{E}}) + {K'}\mu_{\textsc{S}}^{(2)} \leq {K'}\mu_{\textsc{E},\epsilon}^{(2)} - \epsilon$. Now, first, suppose $r(\mathcal{X}^\textsc{E}) \geq 0$. In this case, ${K'} \geq 1$, that is, player 2 has a profitable deviation and so punishment is necessary for $\epsilon$-enforceability.  Then, further, $(\tau_t - t - {K'})\mu_{\textsc{S}}^{(2)} \leq (\tau_t - t - {K'})\mu_{\textsc{E},\epsilon}^{(2)} - \left(\frac{\tau_t - t - {K'}}{{K'}}\right)\epsilon$. Given that for any $t \in \mathcal{T}$ we have $A_t^{(1)} = a_{\textsc{E}}^{(1)}(Y_{t}^{(1)})$ and $A_t^{(2)} \neq a_{\textsc{E}}^{(2)}(Y_{t}^{(1)})$:
\begin{align*}
    \sum_{t\in \mathcal{T}} \left(R_t^{(2)} + \sum_{t'=t+1}^{\tau_t} R_{t'}^{(2)}\right) &\leq \sum_{t\in \mathcal{T}} \left(r(\mathcal{X}_{\textsc{E}}) + R_{\textsc{E},t}^{(2)} + \sum_{t'=t+1}^{\tau_t} R_{t'}^{(2)}\right) \tag{since $t \in \mathcal{T}$} \\
    &\leq \sum_{t\in \mathcal{T}} \left({K'}\mu_{\textsc{E},\epsilon}^{(2)} - \epsilon - {K'}\mu_{\textsc{S}}^{(2)} + R_{\textsc{E},t}^{(2)} + \sum_{t'=t+1}^{\tau_t} R_{t'}^{(2)}\right) \tag{by enforceability} \\
    &\leq \sum_{t\in \mathcal{T}} \left((\tau_t - t)\mu_{\textsc{E},\epsilon}^{(2)} - \epsilon + R_{\textsc{E},t}^{(2)} - \left(\frac{\tau_t - t - {K'}}{{K'}}\right)\epsilon - (\tau_t - t)\mu_{\textsc{S}}^{(2)} + \sum_{t'=t+1}^{\tau_t} R_{t'}^{(2)}\right) \\
    &= \sum_{t\in \mathcal{T}} \left((\tau_t - t)\mu_{\textsc{E},\epsilon}^{(2)} + R_{\textsc{E},t}^{(2)} - \left(\frac{\tau_t - t}{{K'}}\right)\epsilon - (\tau_t - t)\mu_{\textsc{S}}^{(2)} + \sum_{t'=t+1}^{\tau_t} R_{t'}^{(2)}\right).
\end{align*}
Let $\tau_{\mathcal{T}} := |\mathcal{T}|$ and $\tau_{M} := \sum_{t \in \mathcal{T}} (\tau_t - t)$. Let $E := \mathbb{E}\left(\sum_{t \in \mathcal{T}} \sum_{t'=t+1}^{\tau_t} R_{t'}^{(2)}\right)$. Because $\phi_E$ punishes for $t'$ such that $M_{t'}$ holds, $E \leq \tau_M\mu_{\textsc{S}}^{(2)}$. Then, again by Hoeffding:
\begin{align*}
    P\left( \sum_{t \in \mathcal{T}} R_{\textsc{E},t}^{(2)} \geq \tau_{\mathcal{T}} \left[r_{1}^{(2)} + (r_{0}^{(2)} - r_{1}^{(2)})\left(\alpha_{\textsc{E}} + \sqrt{\frac{\log(T/\delta)}{2\tau_{\mathcal{T}}}}\right)\right]\right) &\leq \frac{\delta}{T}, \\
    P\left( \sum_{t \in \mathcal{T}} \sum_{t'=t+1}^{\tau_t} R_{t'}^{(2)} \geq \tau_M \mu_{\textsc{S}}^{(2)} + \sqrt{\frac{ \tau_M\log(T/\delta)}{2}}\right) &\leq P\left( \sum_{t \in \mathcal{T}} \sum_{t'=t+1}^{\tau_t} R_{t'}^{(2)} \geq E + \sqrt{\frac{\tau_M\log(T/\delta)}{2}}\right) \\
    &\leq \frac{\delta}{T}.
\end{align*}
Then, with probability at least $1-\frac{2\delta}{T}$:
\begin{align*}
    \sum_{t\in \mathcal{T}} \left(R_t^{(2)} + \sum_{t'=t+1}^{\tau_t} R_{t'}^{(2)}\right) &\leq \sum_{t\in \mathcal{T}} \left((\tau_t - t)\mu_{\textsc{E},\epsilon}^{(2)} - \left(\frac{\tau_t - t}{{K'}}\right)\epsilon - (\tau_t - t)\mu_{\textsc{S}}^{(2)}\right) \\
    &+ \tau_{\mathcal{T}}\mu_{\textsc{E},\epsilon}^{(2)} + \tau_{\mathcal{T}}(r_{0}^{(2)} - r_{1}^{(2)}) \sqrt{\frac{\log(T/\delta)}{2\tau_{\mathcal{T}}}} + \tau_M \mu_{\textsc{S}}^{(2)} + \sqrt{\frac{ \tau_M\log(T/\delta)}{2}} \\
    &= \sum_{t\in \mathcal{T}} \left((\tau_t - t + 1)\mu_{\textsc{E},\epsilon}^{(2)} - \left(\frac{\tau_t - t}{{K'}}\right)\epsilon \right) \\
    &+ (r_{0}^{(2)} - r_{1}^{(2)}) \sqrt{\frac{\tau_{\mathcal{T}}\log(T/\delta)}{2}} + \sqrt{\frac{\tau_M \log(T/\delta)}{2}} \\
    &\leq \sum_{t\in \mathcal{T}}(\tau_t - t + 1) (\mu_{\textsc{E},\epsilon}^{(2)} - \xi(\epsilon, r(\mathcal{X}_{\textsc{E}}))) + 2 \sqrt{\frac{\tau\log(T/\delta)}{2}} \tag{*} \label{r_bound}.
\end{align*}
In the last step we use the fact that $\tau_t - t + 1 \leq 2(\tau_t - t)$.
Second, suppose $-\epsilon  < r(\mathcal{X}_{\textsc{E}}) < 0$, which still guarantees $K' \geq 1$. Then:
\begin{align*}
    \sum_{t\in \mathcal{T}} \left(R_t^{(2)} + \sum_{t'=t+1}^{\tau_t} R_{t'}^{(2)}\right) &\leq \sum_{t\in \mathcal{T}} \left(r(\mathcal{X}_{\textsc{E}}) + R_{\textsc{E},t}^{(2)} + \sum_{t'=t+1}^{\tau_t} R_{t'}^{(2)}\right) \tag{$t \in \mathcal{T}$} \\
    &\leq \sum_{t\in \mathcal{T}} r(\mathcal{X}_{\textsc{E}}) + \sum_{t\in \mathcal{T}} \mu_{\textsc{E},\epsilon}^{(2)} + \sum_{t\in \mathcal{T}} (\tau_t - t) \mu_{\textsc{S}}^{(2)} + 2\sqrt{\frac{\tau \log(T/\delta)}{2}} \tag{Hoeffding} \\
    &\leq \sum_{t\in \mathcal{T}} r(\mathcal{X}_{\textsc{E}}) + \sum_{t\in \mathcal{T}} \mu_{\textsc{E},\epsilon}^{(2)} + \sum_{t\in \mathcal{T}} (\tau_t - t) \left(\mu_{\textsc{E},\epsilon}^{(2)} - \frac{\epsilon + r(\mathcal{X}_{\textsc{E}})}{{K'}}\right) + 2\sqrt{\frac{\tau \log(T/\delta)}{2}} \tag{enforceability} \\
    &= \sum_{t\in \mathcal{T}} r(\mathcal{X}_{\textsc{E}}) + \sum_{t\in \mathcal{T}} (\tau_t - t + 1) \left(\mu_{\textsc{E},\epsilon}^{(2)} - \frac{\tau_t - t}{(\tau_t - t+1){K'}}(\epsilon + r(\mathcal{X}_{\textsc{E}}))\right) + 2\sqrt{\frac{\tau \log(T/\delta)}{2}}.
\end{align*}
But then the inequality (\ref{r_bound}) holds in this case as well, since $\sum_{t\in \mathcal{T}} r(\mathcal{X}_{\textsc{E}}) \leq 0$. Finally, if $r(\mathcal{X}_{\textsc{E}}) \leq -\epsilon$, then by construction of $\phi_E$, there is no punishment, so $\tau_t - t= 0$, and we have:
\begin{align*}
    \sum_{t\in \mathcal{T}} \left(R_t^{(2)} + \sum_{t'=t+1}^{\tau_t} R_{t'}^{(2)}\right) &\leq  \sum_{t\in \mathcal{T}} r(\mathcal{X}_{\textsc{E}}) + \sum_{t\in \mathcal{T}} \mu_{\textsc{E},\epsilon}^{(2)} + \sqrt{\frac{\tau \log(T/\delta)}{2}} \\
    &= \sum_{t\in \mathcal{T}} (\tau_t - t + 1)(\mu_{\textsc{E},\epsilon}^{(2)} + r(\mathcal{X}_{\textsc{E}})) + \sqrt{\frac{\tau \log(T/\delta)}{2}}.
\end{align*}
Since $r(\mathcal{X}_{\textsc{E}}) \leq -\xi(\epsilon, r(\mathcal{X}_{\textsc{E}}))$, again, inequality (\ref{r_bound}) holds. Putting these parts for player 2's regret together, with probability at least $1-\frac{3\delta}{T}$:
\begin{align*}
    \sum_{t={K'}+1}^{\tau}R_t^{(2)} &\leq
    \sum_{t={K'}+1}^{\tau}R_t^{(2)}\mathbb{I}[D_t \cup M_t] + \tau_{+,{K'}}\left(\mu_{\textsc{E},\epsilon}^{(2)} + (r_{0}^{(2)} - r_{1}^{(2)})\sqrt{\frac{\log(T/\delta)}{2\tau_{+,{K'}}}}\right) \\
    &\leq K' + \sum_{t\in \mathcal{T}}(\tau_t - t + 1) (\mu_{\textsc{E},\epsilon}^{(2)} - \xi(\epsilon, r(\mathcal{X}_{\textsc{E}}))) + 2 \sqrt{\frac{\tau\log(T/\delta)}{2}} + \tau_{+,{K'}}\mu_{\textsc{E},\epsilon}^{(2)} + \sqrt{\frac{\tau\log(T/\delta)}{2}} \\
    &\leq K' +  (\mu_{\textsc{E},\epsilon}^{(2)} - \xi(\epsilon, r(\mathcal{X}_{\textsc{E}})))\left(1 + \sum_{t=K'+1}^{\tau}\mathbb{I}[D_t \cup M_t]\right) + \tau_{+,{K'}}\mu_{\textsc{E},\epsilon}^{(2)} + 3\sqrt{\frac{\tau\log(T/\delta)}{2}} \tag{3} \label{eeeesum}.
\end{align*}
Line \ref{eeeesum} follows because $\tau_t - t + 1$ is one more than the number of time steps $\phi_E$ punishes, so $\sum_{t\in \mathcal{T}} (\tau_t - t + 1) \leq 1 + \sum_{t=K'+1}^{\tau}\mathbb{I}[D_t \cup M_t]$. The first claim of the lemma follows by $(\mu_{\textsc{E},\epsilon}^{(2)} - \xi(\epsilon, r(\mathcal{X}_{\textsc{E}})))(1 + \sum_{t=K'+1}^{\tau}\mathbb{I}[D_t \cup M_t]) \leq 1 + (\tau - \tau_{+,K'})\mu_{\textsc{E},\epsilon}^{(2)}$, since this part of the argument does not require that player 2 is a Follower. Thus with probability at least $1 - \frac{3\delta}{T}$:
\begin{align*}
     \sum_{t={K'}+1}^{\tau}(\mu_*^{(2)} - R_t^{(2)}) &\geq \sum_{t={K'}+1}^{\tau}(\mu_{\textsc{E},\epsilon}^{(2)} - R_t^{(2)}) \\
     &\geq \mu_{\textsc{E},\epsilon}^{(2)}\sum_{t={K'}+1}^{\tau}\mathbb{I}[D_t \cup M_t] - {K'} \\
    &- (\mu_{\textsc{E},\epsilon}^{(2)} - \xi(\epsilon, r(\mathcal{X}_{\textsc{E}})))\left(1 + \sum_{t=K'+1}^{\tau}\mathbb{I}[D_t \cup M_t]\right) - 3 \sqrt{\frac{\tau\log(T/\delta)}{2}} \\
    &\geq \xi(\epsilon, r(\mathcal{X}_{\textsc{E}}))\sum_{t={K'}+1}^{\tau}\mathbb{I}[D_t \cup M_t] - {K'} -1  - 3\sqrt{\frac{(\tau-{K'})\log(T/\delta)}{2}}.
\end{align*}
By stipulation, the Follower's regret is $C_1T_0 + C_2DS^{1/2}A^{1/2}\tau^{1/2}(\log(T\tau/\delta))^{1/2}$ with probability at least $1-\frac{\delta}{T}$, where $T_0$ in general depends on the length of \textit{previous} epochs. Therefore, with probability at least $1-\frac{5\delta}{T}$:
\begin{align*}
    &\tau(\mu_{\textsc{E},\epsilon}^{(1)} - \overline{r}^{(1)}_{i,\tau}) \\
    &\qquad \leq \sum_{t=1}^{\tau} \mathbb{I}[D_t \cup M_t] + \sqrt{\frac{\tau_+\log(T/\delta)}{2}} \\
    &\qquad \leq K' + \frac{1}{\xi(\epsilon, r(\mathcal{X}_{\textsc{E}}))}\left(\xi(\epsilon, r(\mathcal{X}_{\textsc{E}}))\sum_{t=K'+1}^{\tau} \mathbb{I}[D_t \cup M_t]\right) + \sqrt{\frac{\tau_+\log(T/\delta)}{2}} \\
    &\qquad \leq K' + \frac{1}{\xi(\epsilon, r(\mathcal{X}_{\textsc{E}}))}\left(\sum_{t=K'+1}^{\tau}(\mu_{\textsc{E},\epsilon}^{(2)} - R_t^{(2)}) + K' + 1 + 3\sqrt{\frac{\tau\log(T/\delta)}{2}}\right) + \sqrt{\frac{\tau\log(T/\delta)}{2}} \\
    &\qquad \leq K' + \frac{1}{\xi(\epsilon, r(\mathcal{X}_{\textsc{E}}))}\left(C_1T_0 + C_2DS^{1/2}A^{1/2}\tau^{1/2}(\log(\tau/\delta))^{1/2} + K' + 1 + 3\sqrt{\frac{\tau\log(T/\delta)}{2}}\right) + \sqrt{\frac{\tau\log(T/\delta)}{2}}, \\
    &\overline{r}^{(1)}_{i,\tau} \geq \mu_{\textsc{E},\epsilon}^{(1)} - \frac{1}{\tau}\left(K' + \frac{C_1T_0}{\xi(\epsilon, r(\mathcal{X}_{\textsc{E}}))} + \frac{K'+1}{\xi(\epsilon, r(\mathcal{X}_{\textsc{E}}))}\right) \\
    &\qquad - \frac{1}{\tau^{1/2}}\left[\frac{C_2DS^{1/2}A^{1/2}(\log(T\tau/\delta))^{1/2}}{\xi(\epsilon, r(\mathcal{X}_{\textsc{E}}))} + \left(\frac{3}{\xi(\epsilon, r(\mathcal{X}_{\textsc{E}}))}+1\right)\sqrt{\frac{\log(T/\delta)}{2}}\right] \\
    &\qquad \geq \mu_{\textsc{E},\epsilon}^{(1)} - \frac{1}{\tau}\left(K' + \frac{C_1T_0}{\xi(\epsilon, r(\mathcal{X}_{\textsc{E}}))} + \frac{K'+1}{\xi(\epsilon, r(\mathcal{X}_{\textsc{E}}))}\right) - \frac{1}{\tau^{1/2}}\left[\frac{C_2\mathcal{R}_{Q}(\tau, \delta/T)}{\tau^{1/2}\xi(\epsilon, r(\mathcal{X}_{\textsc{E}}))} + \left(\frac{3}{\xi(\epsilon, r(\mathcal{X}_{\textsc{E}}))}+1\right)\sqrt{\frac{\log(T/\delta)}{2}}\right].
\end{align*}
Given that $\mathcal{R}_{Q}(\tau, \delta/T) \geq DS^{1/2}A^{1/2}(\log(T\tau/\delta))^{1/2}$. The same argument applies for the case $V^{(2)} = 0$, using the corresponding target actions and weight and considering $\phi_B$.
\end{proof}

\section{Proof of Lemma \ref{conditional_experts}}

\begin{replemma}{conditional_experts}
\textbf{(False Positive and Negative Control of Exploitation Test)} Consider a sequence of $k$ epochs each of length $H$.
Let $m^*_{F}$ or $m^*_{M}$ be, respectively, the index of the \textit{subepoch} within this sequence at the start of which $\phi_F$ or $\phi_M$ switches to punishing with $\phi_E$, if at all (if not, let $m^*_{F}$ or $m^*_{M} = \infty$). Let $\slackexpl \geq \frac{2\mathcal{R}_{Q}(H/2, \delta/T)}{H} + \sqrt{\frac{2S^2A\log(c_0/\delta)}{c_1H}}$, where $c_0, c_1$ are defined as in Theorem 5.1 of \citet{MT05}, and $\slackmaximin \geq \sqrt{\frac{\log(T/\delta)}{2(H/2-K)}} + \sqrt{\frac{64e\log(N_q/\delta^2)}{(1-\lambda)(H/2-K)}}$, where $\lambda$ and $N_q$ are constants with respect to time defined in Lemma \ref{raolemma} (see Appendix).

Then, suppose player 2 is Bounded Memory, and $\phi_F$ is used. If $\mu_*^{(1)} < V^{(1)} - \slackexpl$, then with probability at least $1-\delta$, $m^*_{F} \leq \ceil{\frac{H^{1/2}}{2}}$. If $\mu_*^{(1)} \geq V^{(1)}$, then with probability at most $\frac{kH^{1/2}\delta}{T}$, $m^*_{F} < \infty$. If $\phi_M$ is used, and $\mu_{\textsc{M}}^{(2)} > \mu_{\textsc{E},\epsilon}^{(2)}$, then with probability at least $1-\delta$, $m^*_{M} \leq \ceil{\frac{H^{1/2}}{2}}$.

Suppose player 2 is Adversarial, with a sequence of action distributions $\{\pi^{(2)}_t\}$ such that, for any $M \geq H^{1/2} - K$ and $i$, $\frac{1}{M} \sum_{t=i+1}^{i+M} {\mathbf{v}^{(1)}_{\textsc{M}}}^\intercal \mathbf{R}^{(2)} \pi^{(2)}_t \leq \mu_{\textsc{E},\epsilon}^{(2)} - \slackmaximin$. Then, if $\phi_M$ is used, with probability at most $\frac{kH^{1/2}\delta}{T}$, $m^*_{M} < \infty$.
\end{replemma}

\begin{proof}
First suppose $\mu_*^{(1)} \geq V^{(1)}$. Let $\tau$ be the number of time steps since the current instance of $\phi_F$ was first deployed. By definition of $\mathcal{R}_{Q}$, we have $P(\tau(\mu_*^{(1)} - \overline{r}_{i,\tau})^{(1)} > \mathcal{R}_{Q}(\tau, \delta/T)) \leq \frac{\delta}{T}$. Then, for any of $kH^{1/2}$ subepochs (and corresponding time $\tau$):
\begin{align*}
    &P\left(\overline{r}_{i,\tau}^{(1)} < V^{(1)} - \frac{\mathcal{R}_{Q}(\tau, \delta/T)}{\tau}\right) \\
    &\qquad \leq P\left(\overline{r}_{i,\tau}^{(1)} < \mu_*^{(1)} - \frac{\mathcal{R}_{Q}(\tau, \delta/T)}{\tau}\right) \\
    &\qquad= P(\tau(\mu_*^{(1)} - \overline{r}_{i,\tau}^{(1)}) > \mathcal{R}_{Q}(\tau, \delta/T)) \\
    &\qquad\leq \frac{\delta}{T}.
\end{align*}
Suppose $\mu_*^{(1)} < V^{(1)} - \slackexpl$. Let $\textbf{r} \in \mathbb{R}^{|\mathcal{S}| \times |\mathcal{A}| \times |\mathcal{S}|}$ be the vector such that $\textbf{r}(s,a,s') = \mathbf{R}^{(1)}(s')$, and $t_i + 1$ be the start time of the sequence of epochs.
As in \citet{MT05}, define $\hat{q}_\tau(s,a,s') := \frac{1}{\tau} \sum_{t=t_i+1}^{t_i+\tau} \mathbb{I}[S_t = s, A_t^{(1)} = a, S_{t+1} = s']$ and for any policy $\pi$, given that in a communicating MDP the initial state does not matter, define $q^{\pi}(s,a,s') := \lim_{T \to \infty} \frac{1}{T} \sum_{t=1}^T \mathbb{E}_{\pi}(\mathbb{I}[S_t = s, A_t^{(1)} = a, S_{t+1} = s'] | S_0)$. Then $\overline{r}_{i,\tau}^{(1)} = \textbf{r}^\intercal \hat{q}_\tau$, and the expected average reward of $\pi$ is $\textbf{r}^\intercal q^{\pi}$. Further, as in \citet{MT05}, let $Q$ be the set of vectors $q$ that satisfy:
\begin{align*}
    q(s,a,s') &= P(s'|s,a) \sum_{s''} q(s,a,s'') \tag{for all $s, a, s'$,} \\
    \sum_{s,a} q(s,a,s') &= \sum_{a',s''} q(s',a',s'') \tag{for all $s'$.}
\end{align*}
Then by Proposition 3.2 of \citet{MT05}, since the MDP induced by a Bounded Memory player 2 is communicating, for any $q \in Q$, there exists a stationary policy $\pi$ such that $q = q^\pi$. By construction of $Q$ by a set of linear constraints,
\adigi{may want to flesh this out}
$Q$ is closed, and so there exists a $q^{\prime \pi} \in Q$ such that $||\hat{q}_\tau - q^{\prime \pi}||_2 = \inf_{q \in Q} ||\hat{q}_\tau - q||_2$. By Theorem 5.1 of \citet{MT05}, there exist constants $c_0, c_1 > 0$ such that $P(\inf_{q \in Q} ||\hat{q}_\tau - q||_2 \geq x) \leq c_0\exp(-c_1 x^2 \tau)$. So:
\begin{align*}
    & P\left(\overline{r}_{i,\tau}^{(1)} \geq V^{(1)} - \frac{\mathcal{R}_{Q}(\tau, \delta/T)}{\tau}\right) \\
    &\qquad \leq P\left(\textbf{r}^\intercal \hat{q}_\tau \geq \textbf{r}^\intercal q^{\pi^*} + \slackexpl - \frac{\mathcal{R}_{Q}(\tau, \delta/T)}{\tau}\right) \\
    &\qquad \leq P\left(\textbf{r}^\intercal (\hat{q}_\tau - q^{\prime \pi}) \geq \slackexpl - \frac{\mathcal{R}_{Q}(\tau, \delta/T)}{\tau}\right) \\
    &\qquad \leq P\left(||\textbf{r}||_2 ||\hat{q}_\tau - q^{\prime \pi}||_2 \geq \slackexpl - \frac{\mathcal{R}_{Q}(\tau, \delta/T)}{\tau}\right) \\
    &\qquad \leq P\left( \inf_{q \in Q} ||\hat{q}_\tau - q||_2 \geq \frac{1}{\sqrt{S^2A}}\left(\slackexpl - \frac{\mathcal{R}_{Q}(\tau, \delta/T)}{\tau}\right)\right) \\
    &\qquad \leq c_0\exp\left(-\frac{c_1 \tau}{S^2A}\left(\slackexpl - \frac{\mathcal{R}_{Q}(\tau, \delta/T)}{\tau}\right)^2\right).
\end{align*}
Now, suppose that up until and excluding the $\ceil{\frac{H^{1/2}}{2}}$'th subepoch, $\phi_F$ has not switched to $\phi_E$. At this subepoch:
\begin{align*}
    & P\left(\overline{r}_{i,\tau}^{(1)} \geq V^{(1)} - \frac{\mathcal{R}_{Q}(\tau, \delta/T)}{\tau}\right) \\
    &\qquad \leq c_0\exp\left(-\frac{c_1 H}{2S^2A}\left(\slackexpl - \frac{\mathcal{R}_{Q}(H/2, \delta/T)}{H/2}\right)^2\right) \\
    &\qquad \leq c_0\exp\left(-\frac{c_1 H}{2S^2A} \cdot \frac{2S^2A\log(c_0/\delta)}{c_1H}\right) \\
    &\qquad = \delta.
\end{align*}
Hence, with probability at least $1-\delta$, we have $m^*_{F} \leq \ceil{\frac{H^{1/2}}{2}}$.
\multilinecomment{
and $m^*_{F} = \Omega(kH^{1/2})$ \adigi{wrong, should be $kH^{1/2} - m^*_{F} = \mathcal{O}(1)$}. Then:
\begin{align*}
    \sum_{t=t_i+1}^{t_i+kH} (V^{(1)} - R^{(1)}_t) &= \sum_{m=1}^{m^*_{F} - 1}\sum_{t=t_i+1+(m-1)H^{1/2}}^{t_i+mH^{1/2}} (V^{(1)} - R^{(1)}_t) + \sum_{m=m^*_{F}}^{kH^{1/2}}\sum_{t=t_i+1+(m-1)H^{1/2}}^{t_i+mH^{1/2}} (V^{(1)} - R^{(1)}_t) \\
    &\leq \mathcal{R}_{Q}((m^*_{F}-1)H^{1/2}, \delta/T) + H^{1/2}(kH^{1/2} - m^*_{F} + 1) \\
    &=  \mathcal{R}_{Q}((m^*_{F}-1)H^{1/2}, \delta/T) + \mathcal{O}(H^{1/2}) \\
    &= \mathcal{O}(\mathcal{R}_{Q}(kH, \delta/T)). \tag{because $m^*_{F} \leq kH^{1/2}$}
\end{align*}
}

Suppose player 2 is Bounded Memory, and suppose that up until and excluding the $\ceil{\frac{H^{1/2}}{2}}$'th subepoch, $\phi_M$ has not switched to $\phi_E$. At this subepoch:
\begin{align*}
    &P\left(\overline{r}^{(2)}_{i,\tau} \leq \mu_{\textsc{E},\epsilon}^{(2)} - \slackmaximin + \sqrt{\frac{\log(T/\delta)}{2(H^{1/2}\ceil{\frac{H^{1/2}}{2}}-\transienceN)}}\right) \\
    &\qquad \leq P\left(\overline{r}^{(2)}_{i,\tau} - \mu_{\textsc{M}}^{(2)} \leq - \slackmaximin + \sqrt{\frac{\log(T/\delta)}{2(\frac{H}{2}-\transienceN)}}\right) \\
    &\qquad \leq \sqrt{N_q \exp\left(-\frac{(1-\lambda)(\frac{H}{2}-\transienceN)}{64e}\left(\sqrt{\frac{\log(T/\delta)}{2(\frac{H}{2}-\transienceN)}} - \slackmaximin\right)^2\right)} \tag{Lemma \ref{raolemma}} \\
    &\qquad \leq \sqrt{N_q \exp\left(-\frac{(1-\lambda)(\frac{H}{2}-\transienceN)}{64e} \cdot \frac{64e\log(N_q/\delta^2)}{(1-\lambda)(\frac{H}{2}-\transienceN)}\right)} \\
    &\qquad = \delta.
\end{align*}
If $\mathbf{v}^{(1)}_M$ is deterministic, then Lemma \ref{raolemma} applies exactly. Otherwise, all transient states are those with player 1 action histories that include actions outside the support of $\mathbf{v}^{(1)}_M$, or inconsistent $y_t^{(i)}$. After $\transienceN$ time steps, the state $s_{K+1}$ must not be such a state, and again the restriction of the MDP to recurrent states is irreducible. Hence Lemma \ref{raolemma} also applies in this case. Thus, with probability at least $1-\delta$, we have $m^*_{M} \leq \ceil{\frac{H^{1/2}}{2}}$.

Now suppose player 2 is Adversarial, with expected rewards bounded as described. Let $\pi^{(2)}_t$ be the (arbitrary) distribution vector over actions used by player 2 at time $t$. Then $R^{(2)}_t$ is distributed such that $P(R^{(2)}_t = \mathbf{R}^{(2)}(i,j)) = (\mathbf{v}^{(1)}_M)_i (\pi_t^{(2)})_j$, with expected value ${\mathbf{v}^{(1)}_M}^\intercal \mathbf{R}^{(2)} \pi^{(2)}_t$. For fixed $\mathbf{v}^{(1)}_M$ and $\pi^{(2)}_t$, the random variables $R^{(2)}_{K+1}, \dots, R^{(2)}_
\tau$ are independent. Thus Hoeffding's inequality applies, and we have, for any of $kH^{1/2}$ subepochs (and corresponding time $\tau$):
\begin{align*}
    &P\left(\overline{r}^{(2)}_{i,\tau} > \mu_{\textsc{E},\epsilon}^{(2)} - \slackmaximin + \sqrt{\frac{\log(T/\delta)}{2(\tau-K)}}\right) \\
    &\qquad \leq P\left(\overline{r}^{(2)}_{i,\tau} - \mathbb{E}(\overline{r}^{(2)}_{i,\tau}) \geq \sqrt{\frac{\log(T/\delta)}{2(\tau-K)}}\right) \\
    &\qquad \leq \exp\left(-2(\tau-K)\left(\sqrt{\frac{\log(T/\delta)}{2(\tau-K)}}\right)^2\right) \\
    &\qquad = \frac{\delta}{T}.
\end{align*}
\end{proof}

\section{Proof of Theorem \ref{hedge}}

\begin{reptheorem}{hedge}
Let $\mathcal{C}$ be the set of player 2 algorithms that are any of the following:
\begin{itemize}
    \item Adversarial, with a sequence of action distributions $\{\pi^{(2)}_t\}$ such that $\frac{1}{M} \sum_{t=i+1}^{i+M} {\mathbf{v}^{(1)}_M}^\intercal \mathbf{R}^{(2)} \pi^{(2)}_t \leq \mu_{\textsc{E},\epsilon}^{(2)} - \slackmaximin$ for any $M \geq T^{1/4}$ and $i$,
    \item Follower, with $V^{(2)} \in \{0, \mu_{\textsc{E},\epsilon}^{(2)}\}$, or
    \item Bounded Memory, with 
    $\mu_*^{(1)} \geq V^{(1)}$.
\end{itemize}
Let $y$ and $\slackexpl$ satisfy the conditions of Lemma \ref{conditional_experts}.
Then, with probability at least $1-5\delta$, LAFF satisfies:
\begin{align*}
    \max_{\mathcal{C}} \mathcal{R}(T) &= \mathcal{O}(\mathcal{R}_{Q}(T, \delta/T)).
\end{align*}
Further, with probability at least $1-6\delta$, LAFF is 
$(V^{(1)},\slackexpl)$-non-exploitable 
when there exists an enforceable EBS.
\end{reptheorem}

\begin{proof}
We consider each case in turn, and let $H = \floor{T^{1/2}}$. For each expert index $j=1,...,5$, let $k_j$ be the index of the epoch in which $\phi_j$ is switched to $\phi_{j+1}$, if at all (otherwise, define $k_j = \infty$).

\paragraph{\textbf{Non-exploitative Bounded Memory.}} In general, if the total regret is bounded by the sum of a constant number of consecutive regret terms each bounded by $\mathcal{O}(\mathcal{R}_{Q}(\tau, \delta/T))$, where $\tau$ is the length of time for that regret term, then total regret is  $\mathcal{O}(\mathcal{R}_{Q}(T, \delta/T))$. For brevity, we will only state the proofs of the bounds of these respective terms, also using the fact that with probability at least $1 - \frac{H^{3/4}\delta}{T} \geq 1-\delta$, the exploitation test is negative every time by Lemma \ref{conditional_experts}. For Bounded Memory players, we need to consider the different possible orderings of $\mu_*^{(1)}$ relative to the target values $\{\mu_j\}$.

If $\mu_*^{(1)} \geq \mu_{\textsc{B},\epsilon}^{(1)}$, for each epoch in which $\phi_F$ is used, with probability at least $1-\frac{\delta}{T}$ we have $\tau(\mu_*^{(1)} - \overline{r}^{(1)}_{i,\tau}) \leq \mathcal{R}_{Q}(\tau, \delta/T)$, so $\overline{r}^{(1)}_{i,\tau} \geq \mu_*^{(1)} - \frac{\mathcal{R}_{Q}(\tau, \delta/T)}{\tau} \geq \mu_{\textsc{B},\epsilon}^{(1)} - \mathcal{B}(\tau)$. Thus Optimistic Q-learning is used each epoch, and so with probability at least $1-2\delta$, $\mathcal{R}(T) = \mathcal{O}(\mathcal{R}_{Q}(T, \delta/T))$. Otherwise, if $k_1 = \infty$, this same argument applies.
If $k_1 < \infty$, up to the end of epoch $k_1$, LAFF will have used $\phi_F$ continuously, so with probability at least $1-\frac{\delta}{T}$, the bound $\mathcal{R}_{Q}$ holds for the first $k_1H$ time steps. If $k_2 = \infty$, this implies that $\overline{r}_{i,\tau}^{(1)} \geq \mu_{\textsc{B},\epsilon}^{(1)} - \mathcal{B}(\tau)$ for every epoch $i$ up to the end of the second-to-last epoch. Thus, with probability at least $1-2\delta$, the remaining regret is bounded as $\sum_{t=k_1H+1}^{T} (\mu_*^{(1)} - R^{(1)}_t) \leq \sum_{t=k_1H+1}^{T} (\mu_{\textsc{B},\epsilon}^{(1)} - R^{(1)}_t) \leq H + (T - (k_1+1)H)\mathcal{B}(T - (k_1+1)H)$. Since $\tau\mathcal{B}(\tau) = \mathcal{O}(\mathcal{R}_{Q}(\tau, \delta/T))$, the result follows.
\multilinecomment{
\begin{align*}
    \mathcal{R}(T) &= \sum_{i=1}^{k_1} \sum_{t=(i-1)H + 1}^{iH} (\mu_*^{(1)} - R^{(1)}_t) + \sum_{i=k_1+1}^{T/H} \sum_{t=(i-1)H + 1}^{iH} (\mu_*^{(1)} - R^{(1)}_t) \\
    &\leq \mathcal{R}_{Q}(k_1H, \delta/T) + \sum_{i=k_1+1}^{T/H} \sum_{t=(i-1)H + 1}^{iH} (\mu_{\textsc{B},\epsilon}^{(1)} - R^{(1)}_t) \\
    &\leq \mathcal{R}_{Q}(k_1H, \delta/T) + H + (T - (k_1+1)H)\mathcal{B}(T - (k_1+1)H) \\
    &= \mathcal{O}(\mathcal{R}_{Q}(T, \delta/T)).
\end{align*}
}

If $k_2 < \infty$ and $\mu_*^{(1)} \geq \mu_{\textsc{E},\epsilon}^{(1)}$, then with probability at least $1-\frac{(T/H-k_2)\delta}{T}$, we have $\overline{r}_{i,\tau}^{(1)} \geq \mu_{\textsc{E},\epsilon}^{(1)} - \mathcal{B}(\tau)$ for all epochs $i$ afterwards, so $k_3 = \infty$. Thus, with probability at least $1 - 2\delta$, the first two terms are bounded as in the last case, and the third term is $\mathcal{O}(\mathcal{R}_{Q}(T - k_2H, \delta/T))$.
\multilinecomment{
\begin{align*}
    \mathcal{R}(T) &\leq \mathcal{R}_{Q}(k_1H, \delta/T) + H + (k_2 - k_1 - 1)H\mathcal{B}((k_2 - k_1 - 1)H) + \sum_{i=k_2+1}^{T/H} \sum_{t=(i-1)H + 1}^{iH} (\mu_*^{(1)} - R^{(1)}_t) \\
    &\leq \mathcal{O}(\mathcal{R}_{Q}(k_2H, \delta/T)) + \mathcal{R}_{Q}(T - k_2H, \delta/T) \\
    &= \mathcal{O}(\mathcal{R}_{Q}(T, \delta/T)).
\end{align*}
}
If $\mu_*^{(1)} < \mu_{\textsc{E},\epsilon}^{(1)}$, yet $k_3 = \infty$, LAFF has used $\phi_F$ indefinitely after $k_2$, thus the same bound as directly above holds with probability at least $1-2\delta$. 
If $k_3 < \infty$ but $k_4 = \infty$, after $k_3$ we will have $\overline{r}_{i,\tau}^{(1)} \geq \mu_{\textsc{E},\epsilon}^{(1)} - \mathcal{B}(\tau)$ for all but possibly the last epoch. Hence, with probability at least $1 - 2\delta$, along with the first three terms we have a bound of $H + (T - (k_3 + 1)H)\mathcal{B}(T - (k_3 + 1)H)$ for the last term.
\multilinecomment{
\begin{align*}
    \mathcal{R}(T) &\leq \mathcal{R}_{Q}(k_1H, \delta/T) + H + (k_2 - k_1 - 1)H\mathcal{B}((k_2 - k_1 - 1)H) \\
    &+ \mathcal{R}_{Q}((k_3 - k_2)H, \delta/T) + \sum_{i=k_3+1}^{T/H} \sum_{t=(i-1)H + 1}^{iH} (\mu_*^{(1)} - R^{(1)}_t) \\
    &\leq \mathcal{O}(\mathcal{R}_{Q}(k_3H, \delta/T)) + H + (T - (k_3 + 1))H\mathcal{B}((T - (k_3 + 1))H) \\
    &= \mathcal{O}(\mathcal{R}_{Q}(T, \delta/T)).
\end{align*}
}

Lastly, if $k_4 < \infty$, LAFF will always use $\phi_F$ thereafter as long as we do not have $\overline{r}_{i,\tau}^{(1)} < \mu_{\textsc{S}}^{(1)} - \mathcal{B}(\tau)$. But since $\mu_*^{(1)} \geq \mu_{\textsc{S}}^{(1)}$, with probability at least $1-\frac{\ceil{T/H}\delta}{T}$ that never happens.
So, with probability at least $1-2\delta$, the first four terms are bounded as in the last case and the last term is $\mathcal{O}(\mathcal{R}_{Q}(T-k_4H, \delta/T))$.
\multilinecomment{
\begin{align*}
    \mathcal{R}(T) &\leq \mathcal{R}_{Q}(k_1H, \delta/T) + H + (k_2 - k_1 - 1)H\mathcal{B}((k_2 - k_1 - 1)H) \\
    &+ \mathcal{R}_{Q}((k_3 - k_2)H, \delta/T) + H + (k_4 - k_3 - 1)H\mathcal{B}((k_4 - k_3 - 1)H) + \sum_{i=k_4+1}^{T/H} \sum_{t=(i-1)H + 1}^{iH} (\mu_*^{(1)} - R^{(1)}_t) \\
    &\leq \mathcal{O}(\mathcal{R}_{Q}(k_4H, \delta/T)) + \mathcal{R}_{Q}(T-k_4H, \delta/T) \\
    &= \mathcal{O}(\mathcal{R}_{Q}(T, \delta/T)).
\end{align*}
}

\paragraph{\textbf{Follower, $V^{(2)} = 0$ (Unconditional)}} If $k_1 = \infty$, we always have $\overline{r}_{i,\tau}^{(1)} \geq \mu_{\textsc{B},\epsilon}^{(1)} - \mathcal{B}(\tau)$. Then $\mathcal{R}(T) \leq T\mathcal{B}(T) = \mathcal{O}(\mathcal{R}_{Q}(T, \delta/T))$.
\multilinecomment{
\begin{align*}
    \mathcal{R}(T) &\leq T\mathcal{B}(T) \\
    &= \mathcal{O}(\mathcal{R}_{Q}(T, \delta/T)).
\end{align*}
}
Otherwise, by Lemma \ref{followregret}, we will for each of $\ceil{T/H}-k_1$ epochs have $\overline{r}_{i,\tau}^{(1)} \geq \mu_{\textsc{B},\epsilon}^{(1)} - \mathcal{B}(\tau)$ with probability $1-\frac{5\delta}{T}$ after $k_1$ by using $\phi_B$. So, with probability at least $1-5\delta$, the first term is bounded by $H + (k_1-1)H\mathcal{B}((k_1-1)H) = \mathcal{O}(\mathcal{R}_{Q}(k_1H, \delta/T))$ and the second by $ (T - k_1H)\mathcal{B}(T-k_1H) = \mathcal{O}(\mathcal{R}_{Q}(T - k_1H, \delta/T))$.
\multilinecomment{
\begin{align*}
    \mathcal{R}(T) &\leq H + (k_1-1)H\mathcal{B}((k_1-1)H) + (T - k_1H)\mathcal{B}(T-k_1H) \\
    &= \mathcal{O}(\mathcal{R}_{Q}(T, \delta/T)).
\end{align*}
}

\paragraph{\textbf{Follower, $V^{(2)} = \mu_{\textsc{E},\epsilon}^{(2)}$ (Conditional)}} If $k_1 = \infty$, we always have $\overline{r}_{i,\tau}^{(1)} \geq \mu_{\textsc{B},\epsilon}^{(1)} - \mathcal{B}(\tau) \geq \mu_{\textsc{E},\epsilon}^{(1)} - \mathcal{B}(\tau)$. So the same proof holds as for the first case of the Unconditional Follower. Otherwise, if $k_2 = \infty$, again we always have $\overline{r}_{i,\tau}^{(1)} \geq \mu_{\textsc{E},\epsilon}^{(1)} - \mathcal{B}(\tau)$ after $k_1$, and the same proof holds as for the second case.
If $k_2 < \infty$ but $k_3 = \infty$, then after $k_2$ we will always have $\overline{r}_{i,\tau}^{(1)} \geq \mu_{\textsc{E},\epsilon}^{(1)} - \mathcal{B}(\tau)$, so the third regret term is bounded by $(T - k_2H)\mathcal{B}(T - k_2H) = \mathcal{O}(\mathcal{R}_{Q}(T - k_2H, \delta/T))$ and the result follows.
\multilinecomment{
Then:
\begin{align*}
    \mathcal{R}(T) &\leq H + (k_1-1)H\mathcal{B}((k_1-1)H) + H + (k_2 - k_1 - 1)H\mathcal{B}((k_2 - k_1 - 1)H) \\
    &+ \sum_{i=k_2+1}^{T/H} \sum_{t=(i-1)H + 1}^{iH} (\mu_{\textsc{E},\epsilon}^{(1)} - R^{(1)}_t) \\
    &\leq \mathcal{O}(\mathcal{R}_{Q}(k_2H, \delta/T)) + (T - k_2H)\mathcal{B}(T - k_2H) \\
    &= \mathcal{O}(\mathcal{R}_{Q}(T, \delta/T)).
\end{align*}
}
Finally, if $k_3 < \infty$, by Lemma \ref{followregret}, we will always have $\overline{r}_{i,\tau}^{(1)} \geq \mu_{\textsc{E},\epsilon}^{(1)} - \mathcal{B}(\tau)$ with probability $1-\frac{5\delta}{T}$ after $k_3$ by using $\phi_E$. So with probability at least $1-5\delta$, the result follows by the same logic as the second Unconditional Follower case.
\multilinecomment{
\begin{align*}
    \mathcal{R}(T) &= H + (k_1-1)H\mathcal{B}((k_1-1)H) + H + (k_2 - k_1 - 1)H\mathcal{B}((k_2 - k_1 - 1)H) \\
    &+ H + (k_3 - k_2 - 1)H\mathcal{B}((k_3 - k_2 - 1)H) + \sum_{i=k_3+1}^{T/H} \sum_{t=(i-1)H + 1}^{iH} (\mu_{\textsc{E},\epsilon}^{(1)} - R^{(1)}_t) \\
    &\leq \mathcal{O}(\mathcal{R}_{Q}(k_3H, \delta/T)) + (T - k_3H)\mathcal{B}(T - k_3H) \\
    &= \mathcal{O}(\mathcal{R}_{Q}(T, \delta/T)).
\end{align*}
}

\paragraph{\textbf{Adversarial}} Since $\mu_{\textsc{S}}^{(1)} \leq \mu_{\textsc{E},\epsilon}^{(1)}$, all the arguments for Conditional Follower above go through except we do not have a guarantee that $k_4 = \infty$ with high probability. If $k_4 < \infty$, up to $k_4$ we still have $\overline{r}_{i,\tau}^{(1)} \geq \mu_{\textsc{E},\epsilon}^{(1)} - \mathcal{B}(\tau) \geq \mu_{\textsc{S}}^{(1)} - \mathcal{B}(\tau)$ (except possibly for epochs $k_1$, $k_2$, $k_3$, and $k_4$). Now, if $k_5 = \infty$, then after $k_4$ we always have $\overline{r}_{i,\tau}^{(1)} \geq \mu_{\textsc{S}}^{(1)} - \mathcal{B}(\tau)$. So with probability at least $1-5\delta$, each term $k$ for a length of time $\tau_{k}$ is bounded by $H + (\tau_k - H) \mathcal{B}(\tau_k - H)$, and the result follows.
\multilinecomment{
\begin{align*}
    \mathcal{R}(T) &= H + (k_1-1)H\mathcal{B}((k_1-1)H) + H + (k_2 - k_1 - 1)H\mathcal{B}((k_2 - k_1 - 1)H) \\
    &+ H + (k_3 - k_2 - 1)H\mathcal{B}((k_3 - k_2 - 1)H) + H + (k_4 - k_3 - 1)H\mathcal{B}((k_4 - k_3 - 1)H) \\
    &+ \sum_{i=k_4+1}^{T/H} \sum_{t=(i-1)H + 1}^{iH} (\mu_{\textsc{S}}^{(1)} - R^{(1)}_t)\\
    &\leq \mathcal{O}(\mathcal{R}_{Q}(k_4H, \delta/T)) + (T - k_4H)\mathcal{B}(T - k_4H) \\
    &= \mathcal{O}(\mathcal{R}_{Q}(T, \delta/T)).
\end{align*}
}
If $k_5 < \infty$, with probability at least $1-\frac{H^{3/2}\delta}{T} \geq 1-\delta$ we never switch to $\phi_E$, and instead play the maximin policy for the rest of the game, by Lemma \ref{conditional_experts}. In that case, by Hoeffding, we have $P\left(\sum_{t=k_5H + 1}^{T} R^{(1)}_t \leq (T - k_5H)\mu_{\textsc{S}}^{(1)} - \sqrt{\frac{(T - k_5H)\log(\frac{1}{\delta})}{2}} \right) \leq \delta$,
since the maximin policy guarantees that $\mathbb{E}(R^{(1)}_t) \geq \mu_{\textsc{S}}^{(1)}$. Therefore, with probability at least $1-2\delta$, the same bound for the first five terms applies as in the previous case, and the last term is $\mathcal{O}(\mathcal{R}_{Q}(T - k_5H, \delta/T))$.
\multilinecomment{
\begin{align*}
    \mathcal{R}^{(1)}(T) &= H + (k_1-1)H\mathcal{B}((k_1-1)H) + H + (k_2 - k_1 - 1)H\mathcal{B}((k_2 - k_1 - 1)H) \\
    &+ H + (k_3 - k_2 - 1)H\mathcal{B}((k_3 - k_2 - 1)H) + H + (k_4 - k_3 - 1)H\mathcal{B}((k_4 - k_3 - 1)H) \\
    &+ H + (k_5 - k_4 - 1)H\mathcal{B}((k_5 - k_4 - 1)H) + \sum_{i=k_5+1}^{T/H} \sum_{t=(i-1)H + 1}^{iH} (\mu_{\textsc{S}}^{(1)} - R^{(1)}_t)\\
    &\leq \mathcal{O}(\mathcal{R}_{Q}(k_5H, \delta/T)) + \sqrt{\frac{(T - k_5H)\log(1/\delta)}{2}} \\
    &= \mathcal{O}(\mathcal{R}_{Q}(T, \delta/T)).
\end{align*}
}

\paragraph{\textbf{Exploitative Bounded Memory}} Finally, suppose we have both $\mu_*^{(1)} < V^{(1)} - \slackexpl$ and $\mu_{\textsc{M}}^{(2)} > \mu_{\textsc{E}}^{(2)}$. Suppose $k_1$ is infinite.
By Lemma \ref{conditional_experts}, $m^*_{F} \leq \ceil{\frac{H^{1/2}}{2}}$ with probability at least $1-\delta$. Let $\overline{t} := K'+(m^*_{F}-1)H^{1/2}$. Define player 2's regret over a time interval $\mathcal{R}^{(2)}(a,b) := \sum_{t=a}^{b} (\mu_{\textsc{E},\epsilon}^{(2)} + c - R^{(2)}_t)$. So with probability at least $1-\frac{3\delta}{T}$, by Lemma \ref{followregret}:
\begin{align*}
    \sum_{t=\overline{t}+1}^{T} R^{(2)}_t &\leq K' + 1 + (T-\overline{t})\mu_{\textsc{E},\epsilon}^{(2)} + 3\sqrt{\textstyle{\frac{1}{2}}(T-\overline{t})\log(\frac{T}{\delta})}, \\
    \mathcal{R}^{(2)}(1,T) 
    &\geq -\overline{t} + (T-\overline{t})(\mu_{\textsc{E},\epsilon}^{(2)} + c) - \sum_{t=\overline{t}+1}^{T} R^{(2)}_t \\
    &\geq c(T-\overline{t}) - \overline{t} - K' - 1 - 3\sqrt{\textstyle{\frac{1}{2}}(T-\overline{t})\log(T/\delta)} \\
    &= \Omega(T).
\end{align*}
Note that the last step requires $T - \overline{t} = \Omega(T)$, as proven, because we have $m^*_{F}H^{1/2} \leq \frac{T^{1/2}}{2} + 1$.

If $k_1<\infty$, we still have, by the above argument (replacing $T$ with $k_1H$), $\mathcal{R}^{(2)}(1,k_1H) = \Omega(k_1H)$. If $k_2 = \infty$, the above argument also implies $\mathcal{R}^{(2)}(k_1H+1,T) = \Omega(T - k_1H)$ since $\mu_{\textsc{B},\epsilon}^{(2)} \leq \mu_{\textsc{E},\epsilon}^{(2)}$, and the same Hoeffding and enforceability arguments bound player 2's rewards against $\phi_B$. Thus in this case, player 2's regret is $\Omega(T)$ with probability at least $1-6\delta$.

Next, if $k_2<\infty$ but $k_3 = \infty$, since Lemma \ref{conditional_experts} guarantees $\phi_F$ has switched to $\phi_E$ with high probability before the end of the \textit{first} epoch, then after $k_2$ LAFF always uses $\phi_E$. So, again, player 2's regret is $\Omega(T)$ with probability at least $1-6\delta$. If $k_3<\infty$ but $k_4 = \infty$, again after $k_3$ LAFF always uses $\phi_E$ and the same argument applies. If $k_4 < \infty$ but $k_5 = \infty$, the same argument as for the case of $k_3 = \infty$ applies.

Finally, if $k_5<\infty$, $\mathcal{R}^{(2)}(1,k_5H) = \Omega(k_5H)$, and for the rest of the game, we use $\phi_M$. With probability at least $1-\delta$, Lemma \ref{conditional_experts} gives $m^*_{M} \leq \ceil{\frac{H^{1/2}}{2}}$, so $ \mathcal{R}^{(2)}(k_5H+1,T) = \Omega((T/H - k_5)H)$ by the same argument, and with probability at least $1-6\delta$, player 2's regret is $\Omega(T)$.

Hence, in all cases, an exploitative player 2 has linear regret.

\end{proof}

\section{Proof of Lemma \ref{raolemma}}

\begin{lemma} Consider an episode of length $\tau$ within a repeated game, starting from any state, in which player 1 follows a fixed deterministic policy $\pi^{(1)}_D$ and player 2 is Bounded Memory.
Then after $\transienceN$ steps of this episode, the Markov chain induced by these policies is irreducible, with a state space $\mathcal{S}_0 \subset \mathcal{S}$ and transition probabilities given by the restriction of $P(\mathcal{S}'|\mathcal{S} \times \mathcal{A}^{(1)})$ to $\mathcal{S}_0$. Further, let $q$ be the initial state distribution at time $\transienceN+1$, $\boldsymbol{\pi}$ be the stationary distribution of the induced chain, $E_{\boldsymbol{\pi}}$ be the $|\mathcal{S}_0| \times |\mathcal{S}_0|$ matrix each of whose rows is $\boldsymbol{\pi}$, and $N_q := \mathbb{E}_{\boldsymbol{\pi}}\left(\left(\frac{dq}{d\boldsymbol{\pi}}\right)^2\right)$. (Because after $K$ steps the Markov chain is irreducible, $\boldsymbol{\pi}$ must have positive probability on the initial state, so $N_q$ is finite.) Define $||v||_{L_2(\boldsymbol{\pi})} := \sum_{s} \boldsymbol{\pi}(s) v_s^2$ and $\lambda := \max_{||v||_{L_2(\boldsymbol{\pi})} = 1} ||(P - E_{\boldsymbol{\pi}})v||_{L_2(\boldsymbol{\pi})}$, as in \citet{Rao19}.
Lastly, let $\mu_D^{(2)}$ be the average expected reward to player 2 in the Markov reward process defined by $\pi^{(1)}_D$ and $\pi^{(2)}$. Then, for $x > 0$:
\begin{align*}
    P\left(\overline{r}^{(2)}_{i,\tau} - \mu^{(2)}_D \leq -x\right) &\leq \sqrt{N_q \exp\left(-\frac{(1-\lambda)(\tau-\transienceN)x^2}{64e}\right)}.
\end{align*}
\label{raolemma}
\end{lemma}

\begin{proof}

\multilinecomment{
\adigi{FIX} Because $\mathcal{S}$ is finite, $\mathbb{E}(\tau_{Re}) < \infty$. We therefore have, by Markov's inequality, for $N = \transienceN$:
\begin{align*}
    P(\tau_{\text{Re}} > N) &\leq \frac{\mathbb{E}(\tau_{\text{Re}})}{N} \\
    &= \frac{\delta}{2}.
\end{align*}
}
We will show that $\mathcal{S}_0 := \mathcal{S} \ \backslash \ \mathcal{S}_{\text{Tr}}$ is the state space of the induced Markov chain after $K$ steps, hence that chain is irreducible.

Let the start state be $s_1 := (a_{-K+1}^{(1)},\dots,a_{0}^{(1)},a_{-K+1}^{(2)},\dots,a_{0}^{(2)},$ $y^{(1)}_{-K+1},\dots,y_{1}^{(1)},y^{(2)}_{-K+1},\dots,y_{1}^{(2)})$, starting at $t=1$ for notational convenience. Then $s_{K+1} = (\pi^{(1)}(s_1),\dots,\pi^{(1)}(s_{K}),$ $a_{1}^{(2)},$ $\dots,a_{K}^{(2)},$ $y^{(1)}_{1},\dots,y_{K+1}^{(1)},y^{(2)}_{1},\dots,y_{K+1}^{(2)})$ is the start state of the chain after $K$ steps. Let $\overline{s} := (\overline{a}_{-K+1}^{(1)},\dots,\overline{a}_{0}^{(1)},\overline{a}_{-K+1}^{(2)},\dots,\overline{a}_{0}^{(2)},$ $\overline{y}^{(1)}_{-K+1},\dots,\overline{y}_{1}^{(1)},\overline{y}^{(2)}_{-K+1},\dots,\overline{y}_{1}^{(2)}) \in \mathcal{S}_{\text{Tr}}$. Since $\pi^{(2)}(a|s) > 0$ for all $a,s$, this state can only be transient if at least one of the following holds: 1) $w^{(i)} \in \{0, 1\}$ and $\overline{y}^{(i)}_{-k}$ is inconsistent with $w^{(i)}$ for some $i,k$. Or 2) there is some $\overline{a}_{-k}^{(1)}$ such that $\pi^{(1)}(\overline{a}_{-k}^{(1)}|\underline{s}) = 0$ for any state $\underline{s} \in \mathcal{S}_0$ such that $ \underline{s} = (\underline{a}_{-K+1}^{(1)},\dots,$ $\underline{a}_{0}^{(1)},\underline{a}_{-K+1}^{(2)},\dots,\underline{a}_{0}^{(2)},\underline{y}^{(1)}_{-K+1},\dots,\underline{y}_{1}^{(1)},\underline{y}^{(2)}_{-K+1},\dots,\underline{y}_{1}^{(2)})$ with action history that is inconsistent with the deterministic $\pi^{(1)}$; that is, $(\underline{a}_{-K+2}^{(2)},\dots,\underline{a}_{-k}^{(2)}) = (\overline{a}_{-K+1}^{(2)},\dots,\overline{a}_{-k-1}^{(2)})$ and $(\underline{y}_{-K+2}^{(2)},$ $\dots,\underline{y}_{-k}^{(2)}) = (\overline{y}_{-K+1}^{(2)},\dots,\overline{y}_{-k-1}^{(2)})$. We thus have $s_{K+1} \notin \mathcal{S}_{\text{Tr}}$, as $\pi^{(1)}(s_k)$ and $y_k^{(i)}$ are clearly consistent with the preceding states; that is, $s_{K+1}$ does not satisfy either condition. And any state for which either condition holds is not reachable with positive probability from any state in $\mathcal{S}_0$, including $s_{K+1}$. Therefore all states visited with positive probability in this induced chain after $K$ steps are in $\mathcal{S}_0$.

Next, define $\mathbf{R}^{(i)}(s) := \mathbf{R}^{(i)}(a^{(1)}(s), a^{(2)}(s))$ and $f_t(S_t) := \frac{\mathbf{R}^{(2)}(S_t) - \mathbb{E}_{\boldsymbol{\pi}}(\mathbf{R}^{(2)}(S_t))}{\tau - \transienceN}$ for $t = \transienceN+1,\transienceN+2,\dots,\tau$. Since $|f_t(s)| \leq \frac{1}{\tau - \transienceN}$ for all $s$, and the induced chain is irreducible after $K$ steps, we have by Theorem 1.1 of \citet{Rao19}:
\begin{align*}
    P_{\boldsymbol{\pi}}\left(\sum_{t=\transienceN+1}^\tau f_t(S_t) \leq -x\right) &\leq \exp\left(-\frac{(1-\lambda)(\tau-\transienceN)x^2}{64e}\right).
\end{align*}
Proposition 3.10 of \citet{P18} (applying the Cauchy-Schwarz inequality and change of measure) gives:
\begin{align*}
    P(\overline{r}^{(2)}_{i,\tau} - \mu^{(2)}_D \leq -x) &\leq \sqrt{N_q P_{\boldsymbol{\pi}}\left(\sum_{t=\transienceN+1}^\tau f_t(S_t) \leq -x\right)} \\
    &\leq \sqrt{N_q \exp\left(-\frac{(1-\lambda)(\tau-\transienceN)x^2}{64e}\right)}.
\end{align*}
\end{proof}

\section{Details on Numerical Experiments}\label{app:experiment_details}

\subsection{Algorithms}

The chosen algorithms in Table \ref{tab:hyperparams} were, with one exception, top performers in a recent tournament study \citep{CO18}.
We include FTFT as a Bounded Memory algorithm that, in some games, can have $\mu_*^{(1)} > \mu_{\textsc{B},\epsilon}^{(1)}$ or $\mu_*^{(1)} > \mu_{\textsc{E},\epsilon}^{(1)}$. While rather exploitable, FTFT can avoid cycles of mutual punishment to which Leader strategies are prone, and was highly successful in a Prisoner's Dilemma tournament \citep{SPtournament}.
Although Q-learning was outperformed by model-based RL in \citet{CO18}, we found the opposite trend in preliminary experiments, so we include the former.

\subsection{Games and Hyperparameters}

\begin{table}[ht]
    \centering
    \caption{Details of algorithms used in experiments}
    \begin{tabular}{p{0.22\linewidth}p{0.15\linewidth} p{0.53\linewidth}}
    \toprule
        \textbf{Algorithm} & \textbf{Classification} & \textbf{Description and Parameters}  \\
    \midrule
        Bully \citep{LS01} & Leader & Equivalent to our Bully Leader expert. This corresponds to augmenting the state space used by \citet{LS01}'s Bully algorithm to match that of our experts. \\
        & \\
        Forgiving Generalized Tit-for-Tat (FTFT) \citep{SPtournament} & Leader & Equivalent to our Egalitarian Leader expert, except that past deviations from the EBS are punished only with probability $p=0.2$.\\
        & \\
        M-Qubed \citep{CG10} & Follower & An optimistic SARSA-based algorithm that empirically cooperates in self-play. Parameters are as in \citet{CG10}, with $\zeta = 0.05$. We omit the exploration stopping rule (equation 16) to avoid excessive slowdown of the algorithm; \citet{CG10} find that this omission does not noticeably decrease M-Qubed's performance. \\
        & \\
        $\epsilon$-Greedy Q-Learning \citep{WD92} & Follower & Given a discount factor $\gamma = 0.95$, initializes Q-value estimates as $\frac{1}{1-\gamma}$ (as in M-Qubed), and with probability $1-\frac{1}{10+t/10}$ takes the action with the highest Q-value estimate (else, takes a uniformly random action). Using the standard Q-learning rule, $\alpha = \frac{5}{10+t/100}$. \\
        & \\
        Fictitious Play \citep{fictplay} & Follower & Letting $\hat{p}$ be the empirical frequency vector of player 1's past actions (independent of state), plays $a^* = \argmax_a (\hat{p}^\intercal \mathbf{R}^{(2)})_a$. \\
        & \\
        S++ \citep{C14} & Bounded Memory + Follower & Applies the aspiration learning algorithm S to ``actions'' given by a set of Leader and Follower experts. The Leader targets are given by enforceable action sequences, rather than randomization over bargaining solution actions; accordingly, we compute the experts' policies over the simplified state space given by the past joint action, without including the randomization signals. Parameters are as in \citet{C14}. We also set the initial aspiration level for player $i$ to $\mu_{\textsc{E},\epsilon}^{(i)}$, given this remark from the supplement of \citet{CO18}: ``In later studies, we set the initial aspiration level based on a fair, Pareto optimal, target solution.'' \\
        & \\
        Manipulator \citep{PS05} & Bounded Memory + Follower & For the first $\frac{T}{20}$ time steps, uses the Leader expert of S++ for which the user's reward from the target solution is maximized. If its average rewards drop below $\mu_{\textsc{B},\epsilon}^{(2)} - \epsilon'$ for $\epsilon' = 0.025$ sometime after this phase, with probability $p = 0.00005$ it switches to model-based RL. After $\frac{3T}{10}$ more time steps, if the other player is nonstationary, the highest-performing expert is locked in. Otherwise, model-based RL is tested for another $\frac{T}{20}$ time steps, and locked in if the other player remains stationary; otherwise, the highest-performing expert is locked in. The locked-in expert is temporarily overriden by maximin if rewards drop below $\mu_{\textsc{S}}^{(2)} - \epsilon'$. \\
    \bottomrule
    \end{tabular}
    \label{tab:hyperparams}
\end{table}

The taxonomy of reward families is based only on the ordinal rankings of rewards in each bimatrix. By default, we use the cardinal values of 1, 2, 3, and 4 for each game as in Supplementary Figure 1 of \citet{CO18}, normalized to $[0, 1]$. However, some games use different cardinal values, chosen either to ensure that an enforceable Bully solution (distinct from the EBS) exists for $K=1$ if possible, or to otherwise generate more ``interesting'' reward structures. For example, the Asymmetric Win-Win game is designed such that the security value for player 2 is relatively high; this increases player 2's incentive to play the risk-dominant Nash equilibrium, rather than the reward-dominant one.

Although the slack terms $\mathcal{B}(\tau)$ and $\mathcal{R}_{Q}(\tau, \delta/T)$ of LAFF are sufficient to provide the results in Theorem \ref{hedge}, in practice we found that these are highly conservative. Specifically, prior to experiments involving the games in Table \ref{tab:game_list}, we evaluated LAFF's performance on the following training set of games:

\begin{align*}
    \begin{tabular}{|c|c|}
        \hline
            3/4, 3/4 & 0, 1\\
             \hline
            1, 0 & 1/4, 1/4\\
             \hline
        \end{tabular} \\
    \begin{tabular}{|c|c|}
        \hline
            5/8, 5/8 & 3/8, 1\\
             \hline
            1, 3/8 & 0, 0\\
             \hline
        \end{tabular} \\
    \begin{tabular}{|c|c|}
        \hline
            1, 1/2 & 0, 0\\
             \hline
            0, 0 & 1/5, 1\\
             \hline
        \end{tabular} \\
    \begin{tabular}{|c|c|}
        \hline
            0, 1 & 1, 2/3\\
             \hline
            1/3, 0 & 2/3, 1/3\\
             \hline
        \end{tabular}
\end{align*}

The first two games have the same outcome orderings as Symmetric Inferior and Symmetric Unfair, respectively, but the cardinal values differ.

We found that LAFF generally performed as intended under the following conditions. Let $C_3$ be a factor by which the $\tau^{-1/2}$-order term of $\mathcal{B}(\tau)$ is multiplied, and $C_4$ be a factor by which $\mathcal{R}_{Q}$ is multiplied when performing the hypothesis tests of $\phi_F$ and $\phi_M$. We set $C_1 = 0.05$, $C_3 = 0.005$, and $C_4 = 0.005$.

\begin{table}[ht]
    \centering
    \caption{Game matrices used in experiments}
    \begin{tabular}{ccc}
    \toprule
        \textbf{Reward Family} & \textbf{Symmetric} & \textbf{Asymmetric} \\
        \midrule
        Win-Win & \begin{tabular}{|c|c|}
        \hline
            1, 1 & 0, 2/3\\
             \hline
            2/3, 0 & 1/3, 1/3\\
             \hline
        \end{tabular} & \begin{tabular}{|c|c|}
        \hline
            1, 1 & 0, 5/6\\
             \hline
            1/3, 0 & 2/3, 2/3\\
             \hline
        \end{tabular} \\
        & & \\
        Biased & \begin{tabular}{|c|c|}
        \hline
            1/3, 1/3 & 2/3, 1\\
             \hline
            1, 2/3 & 0, 0\\
             \hline
        \end{tabular} & \begin{tabular}{|c|c|}
        \hline
            2/3, 0 & 0, 1\\
             \hline
            1, 2/3 & 1/3, 1/3\\
             \hline
        \end{tabular} \\
         & & \\
        Second-Best & \begin{tabular}{|c|c|}
        \hline
            1/3, 1/3 & 0, 1\\
             \hline
            1, 0 & 2/3, 2/3\\
             \hline
        \end{tabular} & \begin{tabular}{|c|c|}
        \hline
            1, 1/3 & 1/3, 1\\
             \hline
            0, 0 & 2/3, 2/3\\
             \hline
        \end{tabular} \\
         & & \\
        Unfair & \begin{tabular}{|c|c|}
        \hline
            1/2, 1/2 & 1/4, 1\\
             \hline
            1, 1/4 & 0, 0\\
             \hline
        \end{tabular} & \begin{tabular}{|c|c|}
        \hline
            0, 1 & 3/4, 3/4\\
             \hline
            1, 1/4 & 1/4, 0\\
             \hline
        \end{tabular} \\
         & & \\
        Inferior & \begin{tabular}{|c|c|}
        \hline
            4/5, 4/5 & 0, 1\\
             \hline
            1, 0 & 1/5, 1/5 \\
             \hline
        \end{tabular} & \begin{tabular}{|c|c|}
        \hline
            1, 3/4 & 0, 1\\
             \hline
            3/4, 0 & 1/4, 1/4\\
             \hline
        \end{tabular} \\
         & & \\
        Cyclic &  & \begin{tabular}{|c|c|}
        \hline
            0, 1 & 3/4, 3/4\\
             \hline
            1, 0 & 1/4, 1/4\\
             \hline
        \end{tabular} \\
    \bottomrule
    \end{tabular}
    \label{tab:game_list}
\end{table}

\subsection{Round Robin and Replicator Dynamic}

We set $K = 1$ and $\epsilon = 0.05$. Each pair of algorithms plays 50 trials of each of the 11 games, for $T = 2 \cdot 10^5$ rounds each trial. For symmetric games, the order of the players does not matter. Thus, for a pair of algorithms indexed by $i,j$ such that $i < j$, we record the results for the case of player 1 as $i$ and player 2 as $j$, and for the reversed case we copy these results. The rewards of each algorithm pair are averaged over the 50 trials, and over the set of games, providing a bimatrix of empirical rewards in the \textit{learning game} between algorithms.

A single trial of the replicator dynamic experiment proceeds as follows. With $J$ algorithms, we initialize a uniform population distribution $\textbf{p} = \frac{1}{J} \mathbf{1}$. For each of $N$ generations, the bimatrix $(M_1^{(k,g)}, M_2^{(k,g)})$ of average rewards in algorithm pairings from one of the 50 trials (indexed by $k$) was drawn with replacement for each game $g$. To compute an algorithm $i$'s performance against each algorithm, we take the elementwise minimum over the bimatrices: $\textbf{r}^{(g)}_{i,n} := \min\{M_1^{(k,g)}(i,:), {M_2^{(k,g)}}^\intercal(i,:)\}$. The motivation for this is that, for asymmetric games,
we assume a given algorithm may find itself as player 1 or player 2 in a game, but with unknown probability, and so we take the minimum to account for this indexical uncertainty. This matches the motivation for the EBS, as opposed to bargaining solutions where one self-copy is bullied by the other. Thus, the minimum metric incentivizes cooperation between self-copies. Each algorithm's \textit{fitness} in generation $n$ is:
\begin{align*}
    f_{i,n} &:= \left(\frac{1}{G} \sum_{g=1}^G \textbf{r}^{(g)}_{i,n}\right)^\intercal \textbf{p}.
\end{align*}
That is, an algorithm's fitness is a weighted average of its performance against the population of algorithms. Letting $\textbf{f}_n$ be the vector of $f_{i,n}$, $\overline{f} := \frac{1}{J}\mathbf{1} ^\intercal \textbf{f}_n$, and $\odot$ be the elementwise product, the replicator dynamic update rule to the next generation is:
\begin{align*}
    \textbf{p} &\leftarrow \textbf{p} \odot ((1 - \overline{f} )\mathbf{1} + \textbf{f}_n).
\end{align*}
We take 1000 repetitions of these trials, and compute averages and standard deviations over these trajectories. We take the average over full replicator dynamic trials rather than compute fitness in each generation by averages over the 50 trials, because of the consideration of bullying self-copies discussed above. That is, in pairings of an algorithm like Manipulator with itself, averaging over trials would mask the bimodal distribution of rewards each self-copy receives depending on whether it is the Leader or Follower in one trial of a game. Our simulation is therefore based on a more appropriate model of the algorithm users' strategic incentives.

\end{document}